\documentclass[
 aip,
 amsmath,amssymb,
 reprint,%
]{revtex4-2}

\usepackage{graphicx}
\usepackage{subcaption}
\usepackage{dcolumn}
\usepackage{bm}

\usepackage[utf8]{inputenc}
\usepackage[T1]{fontenc}
\usepackage{mathptmx}
\usepackage{etoolbox}
\usepackage{orcidlink}

\makeatletter
\def\@email#1#2{%
 \endgroup
 \patchcmd{\titleblock@produce}
  {\frontmatter@RRAPformat}
  {\frontmatter@RRAPformat{\produce@RRAP{*#1\href{mailto:#2}{#2}}}\frontmatter@RRAPformat}
  {}{}
}%
\makeatother
\begin{document}

\preprint{AIP/123-QED}

\title[Quantum Signatures of Strange Attractors]{Quantum Signatures of Strange Attractors}
\author{Bence Dárdai\,\orcidlink{0009-0007-4034-9310}}
\affiliation{Semilab Ltd., Prielle Kornélia Str. 2/A, H-1117 Budapest, Hungary}

\author{Gábor Vattay\,\orcidlink{0000-0002-0919-9429}}%
 \email{gabor.vattay@ttk.elte.hu}
\affiliation{ Institute of Physics and Astronomy, Eötvös Loránd University, Egyetem tér 1-3., H-1053 Budapest, Hungary
}%

\date{\today}

\begin{abstract}
In classical mechanics, driven systems with dissipation often exhibit complex, fractal dynamics known as strange attractors. This paper addresses the fundamental question of how such structures manifest in the quantum realm. We investigate the quantum Duffing oscillator, a paradigmatic chaotic system, using the Caldirola-Kanai (CK) framework, where dissipation is integrated directly into a time-dependent Hamiltonian. By employing the Husimi distribution to represent the quantum state in phase space, we present the first visualization of a quantum strange attractor within this model. Our simulations demonstrate how an initially simple Gaussian wave packet is stretched, folded, and sculpted by the interplay of chaotic dynamics and energy loss, causing it to localize onto a structure that beautifully mirrors the classical attractor. This quantum "photograph" is inherently smoothed, blurring the infinitely fine fractal details of its classical counterpart as a direct consequence of the uncertainty principle. We supplement this analysis by examining the out-of-time-ordered correlator (OTOC), which shows that stronger dissipation clarifies the exponential growth associated with the classical Lyapunov exponent, thereby confirming the model's semiclassical behavior. This work offers a compelling geometric perspective on open chaotic quantum systems and sheds new light on the quantum-classical transition.
\end{abstract}

\maketitle

\begin{quotation}
In the classical world, systems that are simultaneously pushed by an external force and slowed by dissipation, like friction, often settle into complex, fractal patterns known as strange attractors. But what happens when such a system is governed by the laws of quantum mechanics? This paper explores that question by investigating the quantum version of the Duffing oscillator, a textbook example of a system that exhibits chaos. Using the Caldirola-Kanai framework - an approach where dissipation is woven directly into the system's core equations - we visualize the quantum state in phase space using the Husimi distribution. Our central result is the first-ever visualization of a quantum strange attractor within this framework. We demonstrate how an initially simple quantum wave packet is stretched, folded, and ultimately sculpted by the competing forces of chaotic dynamics and energy loss, causing it to localize onto a structure that beautifully mirrors its classical counterpart. This quantum "photograph" of the attractor is inherently smoothed out, blurring the infinitely fine fractal details of the classical version due to the uncertainty principle. This work provides a new, geometric perspective on the behavior of chaotic quantum systems that are open to their environment, shedding light on the quantum-classical transition.\cite{kapitaniak1993controlling}
\end{quotation}

\section{Introduction}

In classical mechanics, dissipative, driven, nonlinear systems exhibit a rich spectrum of dynamical behaviors, including bifurcations and, under suitable driving and dissipation, strange attractors.  In the quantum domain, however, the interplay of nonlinearity, external driving, and loss presents formidable challenges.  Most treatments to date have relied on open‐system formalisms\cite{breuer2002theory} either solving a Lindblad master equation for the reduced density matrix, thus introducing phenomenological dissipators that enforce complete positivity and capture environmental decoherence or invoking quantum Langevin equations derived from Caldeira–Leggett type system–bath Hamiltonians\cite{caldeira1981influence,ferialdi2017dissipation}, where both friction and stochastic fluctuations appear explicitly.  While these approaches provide powerful tools for studying decoherence and energy relaxation, they often obscure the underlying phase‐space structure that governs the emergence of classical‐like chaos and localization phenomena.

The Caldirola–Kanai (CK) framework\cite{caldirola1941forze,10.1143/ptp/3.4.440} provides an alternative approach in which dissipation is incorporated directly into a time-dependent Hamiltonian for the system alone, without an explicit bath or noise term and introduced initially to model purely dissipative quantum motion—where friction leads to irreversible energy decay but no thermal fluctuations—the CK model has been explored in a variety of contexts, ranging from linear potentials and the harmonic oscillator to free-particle localization and two wavepacket interference\cite{SANZ20141}.  

In this work, we extend the CK approach to driven, dissipative mechanical systems and present, for the first time, phase‐space portraits of its strange attractor in terms of the Husimi distributions\cite{takahashi1985chaos}. We demonstrate our approach on the Duffing equation\cite{kovacic2011duffing, ueda1991survey}. By tracing their evolution under nonlinear driving and damping, we can visualize how dissipation sculpts the Husimi function into stretched, filamentary structures that mirror classical chaotic saddles. The Husimi smoothing\cite{takahashi1985chaos} is then applied to reveal the gradual loss of coherence along unstable manifolds.  In doing so, we provide a complementary, geometric viewpoint to more familiar Lindblad- and Langevin-based analyses\cite{everitt2005signatures,everitt2011quantum,spiller1994emergence,ralph2017observing}, and the non-Hermitian approach of Iomin and Zaslavsky\cite{iomin2003breaking}, highlighting the distinct signatures of pure dissipation in the quantum-chaotic phase space.

The central aim of this study is to characterize the phase-space morphology of dissipative quantum chaos, specifically addressing how the fractal geometry of classical strange attractors survives the smoothing effects of the uncertainty principle. By utilizing the Caldirola-Kanai Hamiltonian, we circumvent the phenomenological decoherence inherent in Lindblad formalisms to isolate the deterministic role of friction in quantum state evolution. The significance of this investigation for nonlinear dynamics lies in establishing that dissipation acts as a requisite localizing force, collapsing the wavefunction onto the classical invariant set. Our analysis of the out-of-time-ordered correlator (OTOC) provides a quantitative criterion for the quantum-classical transition, demonstrating that the recovery of the classical Lyapunov exponent is strictly dependent on the dissipation rate driving the system toward the semiclassical limit.

The paper is organized as follows. In Section II, we provide a brief overview of the Caldirola-Kanai (CK) method for quantizing dissipative systems. Section III explores the semiclassical interpretation of this framework, demonstrating how dissipation gives rise to an effective time-dependent Planck's constant, which naturally drives the system toward a classical limit. We then connect classical and quantum chaos in Section IV by discussing the classical Lyapunov exponent and its quantum analog, the out-of-time-ordered correlator (OTOC).
Section V adapts the definition of the Husimi function to the CK formalism, serving as our primary tool for visualizing phase space. The Duffing equation, the paradigmatic model used in our investigation, is introduced in Section VI. Our numerical approach, based on a Trotter-Suzuki split-operator method, is detailed in Section VII. Finally, Section VIII presents the core results, where we visually compare classical Poincaré maps with quantum Husimi distributions across four distinct dynamical regimes: harmonic, non-chaotic dissipative, conservative chaotic, and chaotic dissipative. This culminates in the first visualization of a quantum strange attractor within the CK framework. In Section IX, we further analyze the semiclassical behavior of the out-of-time-ordered correlator, showing how dissipation reveals a clearer connection to the classical Lyapunov exponent. Finally, we summarize our work and its implications in the Conclusions of Section X.

\section{The Caldirola-Kanai method}

In 1941, Caldirola\cite{caldirola1941forze} demonstrated and later Kanai\cite{10.1143/ptp/3.4.440} discovered that in dissipative mechanical systems — where dissipation is proportional to velocity — one can formulate a time-dependent Hamiltonian. This Hamiltonian can subsequently be quantized. The major advantage of this approach is that it allows for the quantization of dissipative systems. Consequently, we can analyze the wave functions associated with strange attractors in these systems using methods developed for conservative systems. Since friction acts as an interaction between a conservative Hamiltonian system and its dissipative environment, this approach also provides new insights into decoherence and the quantum-classical transition.
 
 Caldirola explored an extension of the Lagrange formalism in classical mechanics by rescaling the Lagrangian with a time-dependent scalar function \( a(t) \):
\[
{\cal L}({\bf x},\dot{\bf x},t)=\frac{1}{a(t)}\left(\frac{1}{2}m\dot{\bf x}^2-V({\bf x},t)\right),
\]
where we can initially set \( a(0)=1 \).
The Euler-Lagrange equations naturally yield the dissipative Newtonian equations of motion:
\begin{equation}
m\ddot{\bf x}+\delta(t)m\dot{\bf x}+\nabla V({\bf x},t)=0,\label{Newton}
\end{equation}
where \( \delta(t)=-\frac{\dot{a}(t)}{a(t)} \). For a constant dissipation rate \( \delta \), the scaling function is given by \( a(t)=e^{-\delta t} \).
The canonical momentum is expressed as the rescaled mechanical momentum:
\[
{\bf p}=\frac{\partial {\cal L}}{\partial \dot{\bf x}}=\frac{m\dot{\bf x}}{a(t)}.
\]
The Hamiltonian can be represented as:
\[
{\cal H}=\dot{\bf x}{\bf p}-{\cal L}=a(t)\frac{\bf p^2}{2m}+\frac{1}{a(t)}V({\bf x},t).
\]
In this formulation, the mechanical energy of the Newtonian system — comprising the sum of kinetic and potential energies — can also be expressed in terms of the canonical momentum:
\[
E=\frac{m\dot{\bf x}^2}{2}+V({\bf x},t)=a^2(t)\frac{{\bf p}^2}{2m}+V({\bf x},t).
\]
Thus, the mechanical energy \( E \) and the Hamiltonian \( {\cal H} \) are no longer equal, but they are related by the equation:
\[
E=a(t) \cdot {\cal H}.
\]

The Hamiltonian can be quantized using the canonical commutators $[\hat{x}_j,\hat{p}_i]=i\hbar \delta_{ij}$. The Schrödinger equation then becomes
\begin{equation}
    i\hbar\partial_t |\psi\rangle=\left[a(t)\frac{\bf \hat{p}^2}{2m}+\frac{1}{a(t)}V({\bf \hat{x}},t)\right]|\psi\rangle .
\end{equation}

\section{Semiclassification \label{section:semi}}

We can observe that by multiplying both sides by \(a(t)\) and introducing the time-dependent reduced Planck's constant \(\hbar(t) = \hbar \cdot a(t)\), we obtain the following equation:
\[
i\hbar(t) \partial_t |\psi\rangle = \left[\frac{\hat{\mathbf{P}}^2}{2m} + V(\hat{\mathbf{x}}, t)\right] |\psi\rangle,
\]
where the new momentum operator \(\hat{\mathbf{P}} = a(t) \cdot \hat{\mathbf{p}}\)  satisfies the new canonical commutation relation \([\hat{{x}_i}, \hat{P}_j] = i\hbar(t) \delta_{ij}\).
We can interpret this equation as describing a fixed Hamiltonian operator given by 
\[
\hat{H} = \frac{\hat{\mathbf{P}}^2}{2m} + V(\hat{\mathbf{x}}, t),
\]
in which Planck's constant exhibits time dependence during the evolution of the system.
In the dissipative case, where \(\delta(t) > 0\), the function \(a(t)\) approaches zero, which implies that the effective Planck constant also approaches zero \(\hbar(t) \rightarrow 0\) as time goes on. In other words, the system naturally tends toward the semiclassical limit over time, resulting in increasingly classical behavior. In the new variables, the effects of dissipation become evident in the Ehrenfest equations
\begin{eqnarray}
    \frac{d}{dt}\langle \psi|{\bf x}|\psi \rangle&=&\frac{1}{m}\langle \psi| {\bf P}|\psi \rangle, \\ 
    \frac{d}{dt}\langle \psi| {\bf P}|\psi \rangle&=&-\delta(t)\langle \psi|{\bf P}|\psi \rangle-\langle \psi|\nabla V({\bf x},t) |\psi\rangle,\nonumber
\end{eqnarray}
where we accounted for the fact that \(\langle \psi|\dot{\mathbf{P}}|\psi \rangle = -\delta(t) \langle \psi| \mathbf{P}|\psi \rangle\) due to the time dependence of the momentum operator. Another significant aspect of this formalism is that the expectation value of the Hamiltonian corresponds to the mechanical energy of the system, represented by \(E = \langle \psi |\hat{H}|\psi\rangle\).

We can also develop the Madelung fluid equations, which provide a hydrodynamic interpretation of the time-dependent Schrödinger equation in the case of the CK. This formulation is achieved by expressing the complex wave function, $\psi(\mathbf{x}, t)$, in its polar form, $\psi = \sqrt{\rho} e^{iS/\hbar(t)}$. The novel feature of the CK approach is that $\hbar(t)$ is time-dependent and decreases over time.
In this representation, $\rho(\mathbf{r}, t)$ is the probability density of the particle, and $S(\mathbf{x}, t)$ is the phase, which corresponds to the action in the Hamilton-Jacobi formalism. By substituting this form into the CK Schrödinger equation and separating the real and imaginary parts, two distinct, real-valued equations emerge. The first is a continuity equation for the probability density, which is analogous to the conservation of mass in a classical fluid:
$$
\frac{\partial \rho}{\partial t} + \nabla \cdot \left(\rho \frac{\nabla S}{m}\right) = 0
$$
Here, the term $\mathbf{v} = \frac{\nabla S}{m}$ is interpreted as the velocity field of the quantum "fluid". The second equation is a quantum Hamilton-Jacobi equation, which governs the evolution of the phase $S$:
$$
\frac{\partial S}{\partial t} + \delta(t)S+\frac{(\nabla S)^2}{2m} + V + Q = 0
$$
This equation is the classical Hamilton-Jacobi equation with the term $\delta(t)S$ describing dissipation with the inclusion of the term $Q$, known as the {\em quantum potential}, defined as:
$$
Q = -\frac{\hbar^2(t)}{2m} \frac{\nabla^2 \sqrt{\rho}}{\sqrt{\rho}},
$$
which now contains the time dependent prefactor $\hbar^2(t)$.
The quantum potential is responsible for all non-classical behaviors, such as interference and tunneling, which distinguish quantum fluids from classical fluids. This reformulation presents quantum mechanics using the language of fluid dynamics, where the probability density behaves like the density of a fluid. This fluid moves with a velocity determined by the phase of the wave function and is influenced by both classical and quantum potentials. A key aspect of the CK approach is that the quantum potential decreases over time, leading to a gradual weakening of quantum effects as time progresses, and the behavior of the wave function becomes effectively semiclassical\cite{cvitanovic2005chaos}.

\section{Classical Lyapunov exponent and quantum OTOC}

In dissipative mechanical systems subject to friction and energy loss, trajectories are funneled toward low-dimensional attractors that may exhibit richly chaotic dynamics, despite being low-dimensional. Lyapunov exponents provide a quantitative link between the microscopic mechanisms of dissipation and the emergent phase-space organization, revealing how exponential stretching along unstable manifolds is counterbalanced by volume contraction induced by damping and external driving. For such systems, a positive maximal Lyapunov exponent signals sustained chaotic motion on a strange attractor bounded by the system’s dissipative constraints\cite{zeni1995lyapunov}.

For a classical trajectory $x(t)=x(x(0),{p}(0),t)$ with initial $x(0)$ and ${p}(0)$, the sensitivity to the initial conditions can be characterized by the Lyapunov exponent $\lambda$, the rate of divergence of nearby trajectories
\begin{equation}
    \frac{\partial x(t)}{\partial x(0)}\sim e^{\lambda t},
\end{equation}
for sufficiently long times $t$ for trajectories converging to an attractor.
In classical mechanics, this can be written in the form of a Poisson bracket\cite{garcia2022out}
\begin{equation}
    B(t)=\left\{x(t),p(0)\right\}=\frac{\partial x(t)}{\partial x(0)}\underbrace{\frac{\partial p(0)}{\partial p(0)}}_{1}-\frac{\partial x(t)}{\partial p(0)}\underbrace{\frac{\partial p(0)}{\partial x(0)}}_{0}=\frac{\partial x(t)}{\partial x(0)}.
\end{equation}
Taking the first and second time derivatives of the Poisson bracket and using the Hamiltonian equations of motion $\dot{x}=a(t)p/m$ and $\dot{p}=-V'(x,t)/a(t)$ yields
\begin{eqnarray}
    \dot{B}(t)&=&\frac{a(t)}{m}\left\{p(t),p(0)\right\} \nonumber \\
    \ddot{B}(t)&=&\frac{\dot{a}(t)}{m}\left\{p(t),p(0)\right\}-\frac{1}{m}V''(x,t)\left\{x(t),p(0)\right\},
\end{eqnarray}
which leads to the linearized Newtonian  equations of motion describing the evolution of small perturbations near a trajectory $x(t)$:
\begin{equation}
    \ddot{B}(t)+\delta(t)\dot{B}(t)+\frac{1}{m}V''(x(t),t)B(t)=0.\label{eq:stability}
\end{equation}
The Lyapunov exponent is the long-time limit of the growth rate of the Poisson bracket
\begin{equation}
    \lambda=\lim_{t\rightarrow\infty}\frac{1}{t}\ln |B(t)|.\label{eq:lyapunov}
\end{equation}

The Poisson bracket formulation allows us to generalize this formalism for the quantum
case using the commutator $\left[ \hat{x}(t),\hat{p}(0)\right]$,
where we use the Heisenberg representation of operators $\hat{x}(t)$ and $\hat{p}(0)$.
The out-of-time-ordered correlator\cite{garcia2022out,michel2024quasiclassical} (OTOC) provides a quantum definition of the Lyapunov exponent by probing the sensitivity of operators to initial conditions. For example, one may consider the growth of the commutator between the position operator at time $t$, $x(t)$, and the momentum operator at the initial time, $p(0)$. The OTOC in the Caldirola-Kanai case can be defined as
\begin{equation}
C(t) = -\frac{1}{\hbar^2(t)}\langle [\hat{x}(t),\hat{p}(0)]^{2} \rangle,\label{eq:otoc}
\end{equation}
where the expectation value is taken in some appropriate state. The commutator $[x(t),p(0)]$ captures how much the time-evolved operator $x(t)$ fails to commute with the initial momentum, reflecting the spreading of quantum information. Squaring the commutator is essential: without squaring, the expectation value could vanish due to cancellations between positive and negative contributions, and it would not provide a positive-definite measure of operator growth. In chaotic systems, one typically finds
\begin{equation}
    C(t) \sim e^{2\lambda_{Q} t},\label{eq:lambdaq}
\end{equation}
with $\lambda_{Q}$ identified as the quantum Lyapunov exponent, directly analogous to the exponential separation of trajectories in classical chaos.

\section{The Husimi function}

In this section, we adapt the definition of the Husimi function to the Caldirola-Kanai scenario, employing the time-dependent Planck constant approach developed in the previous section. We focus our analysis on the one-dimensional case.
The Husimi function provides a phase-space representation of quantum states. It is always non-negative and can be interpreted as a probability density. 
A coherent state centered at the phase-space point \((x, P)\) is defined as follows:
\begin{equation}
    \phi_{x,P}(y) = \left(\frac{1}{2\pi \sigma^2}\right)^{1/4} \exp\left[-\frac{(y - x)^2}{4\sigma^2} + i\frac{P y}{\hbar(t)}\right] \label{eq:coherent}
\end{equation}
Here, \(\sigma\) controls the width of the wave packet, and it is important to note that  \(P/\hbar(t) = p/\hbar\).
These states saturate the uncertainty principle, which can be expressed as:
\[
\Delta x \Delta P = \frac{\hbar(t)}{2}, \quad \Delta x = \sigma, \quad \Delta P = \frac{\hbar(t)}{2\sigma}
\]
The Husimi function, denoted as \(Q(x, P)\), is defined as the squared projection of the quantum state \(|\psi\rangle\) onto the coherent state \(\langle \phi_{x,P}|\):
\[
Q(x, P) = \frac{1}{2\pi \hbar(t)} \left| \langle {\phi_{x,P} | \psi} \rangle \right|^2
\]
The factor \((2\pi\hbar(t))^{-1}\) ensures normalization, given by:
\[
\iint Q(x, P) \, dx \, dP = 1
\]
Substituting equation \(\eqref{eq:coherent}\) into the definition of the Husimi function, we have:
\[
\langle \phi_{x,P} | \psi \rangle = \left(\frac{1}{2\pi \sigma^2}\right)^{1/4} \int_{-\infty}^{\infty} \exp\left[-\frac{(y - x)^2}{4\sigma^2} - i\frac{P y}{\hbar(t)}\right] \psi(y) \, dy
\] 

\section{Chaos and Strange Attractors in the Duffing Equation}

The Duffing equation is a paradigmatic nonlinear dynamical system\cite{kovacic2011duffing} exhibiting a rich array of complex behaviors, including periodic oscillations, bifurcations, chaos, and strange attractors. The equation, in its standard forced and damped form, can be written as
\begin{equation}
\label{eq:duffing}
\ddot{x} + \delta \dot{x} + \alpha x + \beta x^3 = \gamma \cos(\omega t),
\end{equation}
where $x(t)$ is the displacement, $\delta$ represents the damping coefficient, $\alpha$ and $\beta$ control the linear and nonlinear stiffness, respectively, and $\gamma \cos(\omega t)$ is an external periodic forcing term. The Duffing system, due to its relative mathematical simplicity yet profound complexity in behavior, remains a critical model in understanding nonlinear dynamics, chaos theory, and associated phenomena such as resonance, bifurcation sequences, and fractal attractors.

One remarkable feature of the Duffing oscillator is the period-doubling cascade to chaos\cite{feigenbaum1978quantitative}. As the forcing amplitude $\gamma$ or frequency $\omega$ is varied, the system undergoes a sequence of period-doubling bifurcations, wherein stable periodic orbits repeatedly bifurcate, doubling their period at each bifurcation. This cascade continues until the system dynamics become chaotic. The presence of chaotic behavior can be quantified through Lyapunov exponents, with positive exponents indicating sensitive dependence on initial conditions, a hallmark of chaos \cite{strogatz2024nonlinear}.

In the chaotic regime, the Duffing oscillator trajectory in phase space no longer settles into simple periodic orbits; rather, it converges onto complex geometric structures known as strange attractors. Strange attractors possess fractal geometry, non-integer dimensions, and demonstrate intricate folding and stretching of trajectories \cite{ott2002chaos}. The fractal nature of these attractors can be visualized through Poincaré sections, revealing detailed, self-similar structures indicative of underlying deterministic chaos.

Moreover, the Duffing equation exemplifies sensitive dependence on initial conditions, whereby infinitesimally small differences in initial states exponentially diverge, resulting in unpredictability on longer time scales. Nevertheless, despite such unpredictability, the strange attractor imposes a deterministic, globally bounded structure on the dynamics.

\section{Numerical solution}

\begin{figure}[h]
    \centering
    \includegraphics[width=0.9\linewidth]{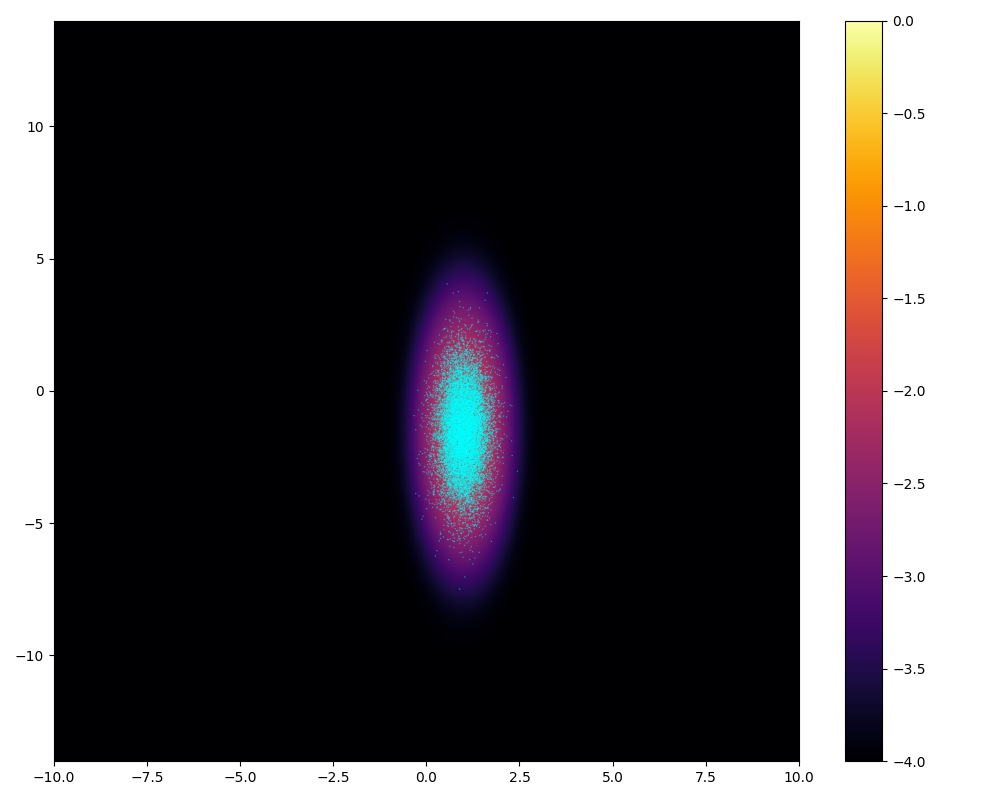}
    \caption{Initial Husimi distribution (\ref{eq:init}) and the distribution of the initial classical trajectories used in the simulations. In our numerical examples, we use the parameters $x_0=+1$, $p_0=-1.5$, $\sigma=0.5$ in units of $\hbar=1$.}
    \label{fig:init}
\end{figure}

The Caldirola-Kanai Hamiltonian of the Duffing Equation can be written as $i \frac{\partial \psi}{\partial t} = \hat{H}(x,t) \psi$, where the Hamiltonian $\hat{H}(x,t)$ is:
$$
\hat{H}(x,t) = \hat{T}(t) + \hat{V}(x,t)
$$
with the kinetic part $\hat{T}(t)$ and potential part $\hat{V}(x,t)$ defined as:
$$
\hat{T}(t) = -\frac{a(t)}{2} \frac{\partial^2}{\partial x^2}
$$
$$
\hat{V}(x,t) = \frac{1}{a(t)} \left( \alpha \frac{x^2}{2} + \beta \frac{x^4}{4} - \gamma x \cos(\omega t) \right)
$$
For a small time step $\Delta t$, the evolution $\psi(t+\Delta t) \approx e^{-i\hat{H}(t_{mid})\Delta t} \psi(t)$ is approximated using a Trotter-Suzuki decomposition. We use $t_{mid} = t_n + \Delta t/2$ for coefficients in the interval $[t_n, t_n+\Delta t]$.

A second-order accurate scheme, which can be computationally efficient regarding FFT operations for time-dependent coefficients, is the V-T-V splitting:
$$
\hat{U}_{step}(t_{mid}, \Delta t) \approx e^{-i\hat{V}_{mid}(x)\frac{\Delta t}{2}} e^{-i\hat{T}_{mid}\Delta t} e^{-i\hat{V}_{mid}(x)\frac{\Delta t}{2}},
$$
where $\hat{T}_{mid} = \hat{T}(t_{mid})$ and $\hat{V}_{mid}(x) = \hat{V}(x,t_{mid})$.
The potential propagator $e^{-i\hat{V}_{mid}(x)\delta t}$ acts as a pointwise multiplication in position space:
$$
\left(e^{-i\hat{V}_{mid}(x)\Delta t} \psi\right)(x) = e^{-i \frac{\Delta t}{a(t_{mid})} \left( \alpha \frac{x^2}{2} + \beta \frac{x^4}{4} - \gamma x \cos(\omega t_{mid}) \right) } \psi(x)
$$

The kinetic propagator $e^{-i\hat{T}_{mid}\Delta t} = \exp\left(i \frac{a(t_{mid})\Delta t}{2} \frac{\partial^2}{\partial x^2}\right)$ is applied in Fourier (momentum) space. If $\mathcal{F}$ denotes the Fourier transform, then:
$$
\left(e^{-i\hat{T}_{mid}\Delta t} \psi\right)(x) = \mathcal{F}^{-1}\left\{ \exp\left(-i \frac{a(t_{mid})\Delta t}{2} k^2\right) \mathcal{F}\{\psi(x)\} \right\}
$$
where $k$ is the wavevector.

The spatial domain for $x$ is discretized into $M=2^N$ points: $x_j = x_{min} + j \Delta x$, for $j = 0, 1, \dots, M-1$. The wavefunction $\psi(x_j, t)$ becomes a vector $\vec{\psi}(t)$ with elements $\psi_j(t)$.

The discrete wavevectors $k_q$ (for $q=0, \dots, M-1$) are used for the Discrete Fourier Transform (DFT). These are typically ordered as $k_q = \frac{2\pi}{M\Delta x} q'$ where $q'$ corresponds to the sequence $[0, 1, \dots, M/2-1, -M/2, \dots, -1]$. The DFT matrix $\mathbf{F}$ and its inverse $\mathbf{F}^{-1}$ are used to transform between position and momentum representations. Using a common physics normalization for unitary transforms:
$(\mathbf{F})_{qj} = \frac{1}{\sqrt{M}} e^{-i k_q x_j}$ and $(\mathbf{F}^{-1})_{jq} = \frac{1}{\sqrt{M}} e^{i k_q x_j}$.
Other normalizations (e.g., factor of $1/M$ on the inverse transform) are also used. The derivation below assumes a $1/M$ factor arises from $\mathbf{F}^{-1} \mathbf{F}$ pair using non-unitary FFT definitions.

\begin{figure*}[t] 
    \centering
    \begin{subfigure}{0.32\textwidth}
        \includegraphics[width=\linewidth]{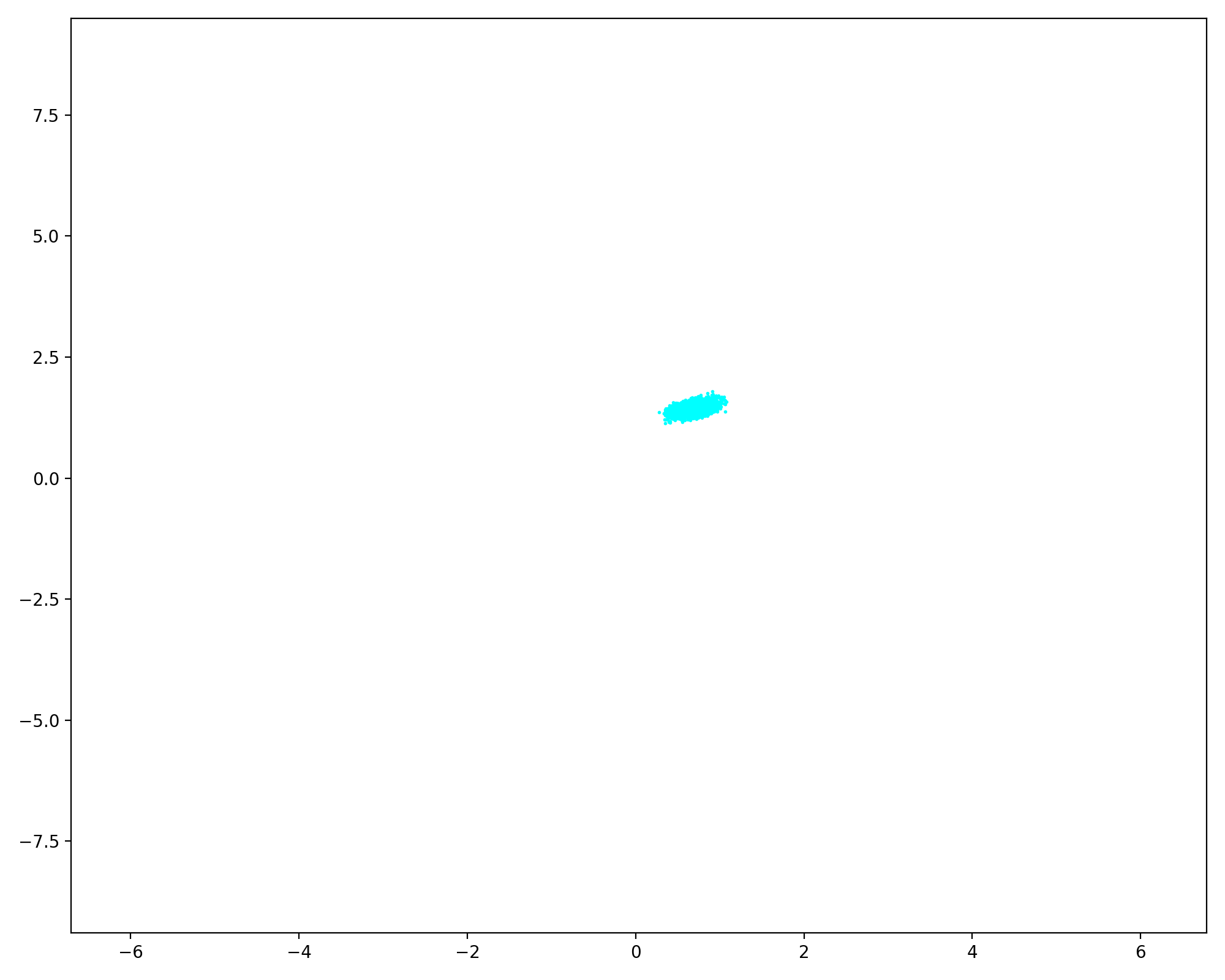}
        \caption{Poincaré return map $(x,P)$}
        \label{fig4:sub1}
    \end{subfigure}%
    \hfill 
    \begin{subfigure}{0.32\textwidth}
        \includegraphics[width=\linewidth]{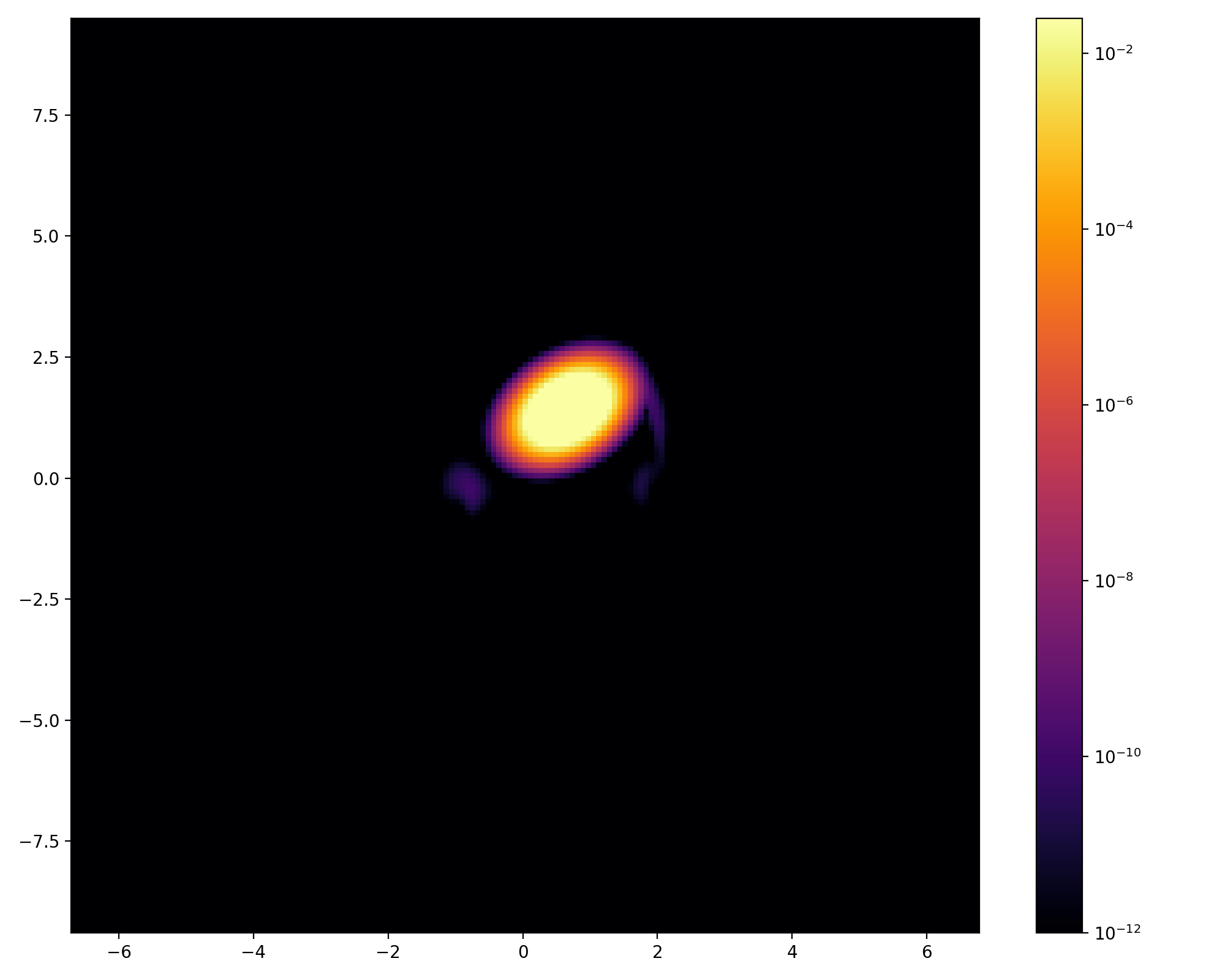}
        \caption{Husimi distribution $\log Q(x,P)$}
        \label{fig4:sub2}
    \end{subfigure}%
    \hfill 
    \begin{subfigure}{0.32\textwidth}
        \includegraphics[width=\linewidth]{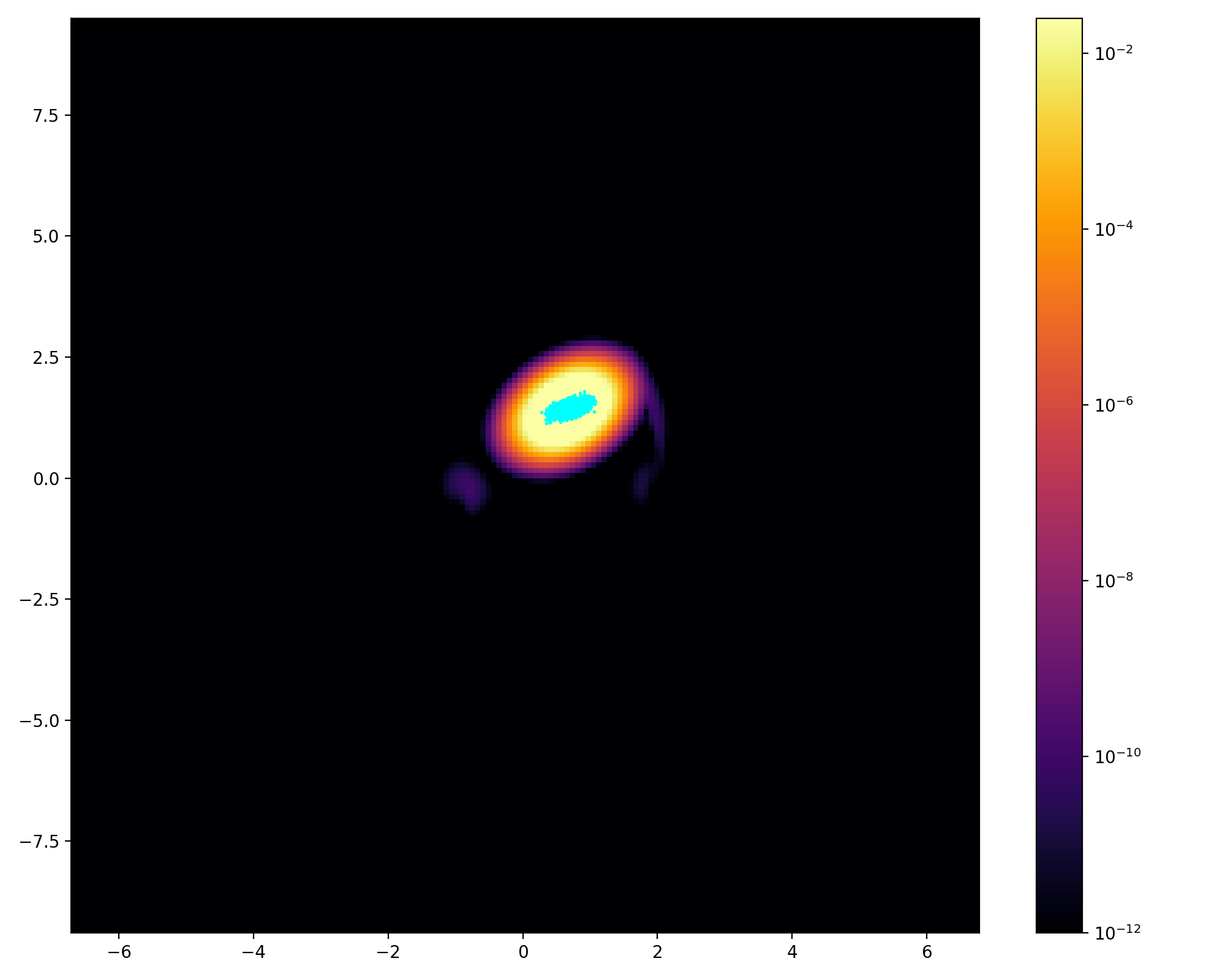}
        \caption{Poincaré + Husimi superimposed }
        \label{fig4:sub3}
    \end{subfigure}
    
    \caption{The harmonic dissipative Duffing equation $\alpha=+1$, $\beta=0$, $\delta=0.1$, $\gamma=2.5$, $\omega=2$. Snapshot of the classical and quantum evolution after $13.37$ cycles $T_{cy}$ of the external forcing.}
    \label{fig4:main}
\end{figure*}

The matrix representation of the one-step V-T-V propagator $\mathbf{U}_{step}$ in the position basis $\{|x_r\rangle\}$ is $\mathbf{U}_{step} = \mathbf{V}_{half} \mathbf{K}_{full} \mathbf{V}_{half}$.
\begin{itemize}
    \item $\mathbf{V}_{half}$ is a diagonal matrix representing $e^{-i\hat{V}_{mid}(x)\frac{\Delta t}{2}}$. Its diagonal elements are:
    $$ D_j(t_{mid}) = e^{-i \frac{\Delta t/2}{a(t_{mid})} \left( \alpha \frac{x_j^2}{2} + \beta \frac{x_j^4}{4} - \gamma x_j \cos(\omega t_{mid}) \right) } $$
    \item $\mathbf{K}_{full}$ is the matrix for $e^{-i\hat{T}_{mid}\Delta t}$, given by $\mathbf{F}^{-1} \mathbf{K}_{diag} \mathbf{F}$, where $\mathbf{K}_{diag}$ is diagonal in momentum space with elements $e^{-i \frac{a(t_{mid})\Delta t}{2} k_q^2}$.
\end{itemize}
The matrix elements $(\mathbf{U}_{step})_{rs} = \langle x_r | \hat{U}_{step} | x_s \rangle$ are:
$$
(\mathbf{U}_{step})_{rs} = D_r(t_{mid})  \frac{1}{M} \sum_{q=0}^{M-1} e^{-i \frac{a(t_{mid})\Delta t}{2} k_q^2} e^{i k_q (x_r - x_s)}  D_s(t_{mid})
$$
This matrix is generally dense.

The choice of the simulation domain is critical and should ensure the wavefunction $\psi(x,t)$ is negligible at the boundaries $x_{min}$ and $x_{max}$ throughout the simulation.
Considerations include the confining nature of the potential $V_{eff}(x,t)$, particularly the $\beta x^4$ term for large $|x|$, the spatial extent of the initial state $\psi(x,0)$, the expected energy range, classical turning points, and dynamical effects like wavepacket motion due to the $\gamma x \cos(\omega t)$ term and natural wavepacket spreading.
The most reliable method is to perform numerical convergence tests by varying the domain size and ensuring results are stable.

If $a(t)$ takes the specific form $a(t) = e^{-\delta t}$, where $\delta$ is a constant rate parameter, then at each time step $t_n \to t_n + \Delta t$, the value used is $a(t_{mid}) = e^{-\delta (t_n + \Delta t/2)}$. The potential propagator elements $D_j$ become:
    $$ D_j = e^{-i e^{\delta t_{mid}} \frac{\Delta t}{2} \left( \alpha \frac{x_j^2}{2} + \beta \frac{x_j^4}{4} - \gamma x_j \cos(\omega t_{mid}) \right)}. $$
The kernel for the kinetic propagator is
    $$ \mathbf{K}_{diag}=e^{-i \frac{e^{-\delta t_{mid}}\Delta t}{2} k_q^2}. $$
The sign of $\delta$ has significant implications. If $\delta > 0$: $a(t)$ decreases, potential strengthens/steepens (scaled by $e^{\delta t_{mid}}$), kinetic term becomes less significant. Tends towards localization. If $\delta < 0$: $a(t)$ increases, potential weakens/flattens, kinetic term becomes more significant. Tends towards delocalization/spreading, requiring careful choice of a sufficiently large domain.

\section{Results}

In the following sections, we show four examples, where we can compare the classical attractor on the Poincaré surface of section and the Husimi distribution in the corresponding phase space $(x, P)$. We start
with a Gaussian wavepacket centered at $(x_0,p_0)$ 
\[
\psi(x) = \bigl(2\pi\sigma^{2}\bigr)^{-1/4}
\exp\!\left[-\frac{(x-x_0)^2}{4\sigma^{2}} + \frac{i p_0 (x-x_0)}{\hbar}\right],
\]
and a coherent state centered at $(q,p)$ with the same width
\[
\phi_{q,P}(x) = \bigl(2\pi\sigma^{2}\bigr)^{-1/4}
\exp\!\left[-\frac{(x-q)^2}{4\sigma^{2}} + \frac{i P (x-q)}{\hbar}\right].
\]
The Husimi function for the Gaussian wavepacket evaluates to
\begin{equation}
Q(x,P) = \frac{1}{2\pi\hbar}\,
\exp\!\left[
-\,\frac{(x-x_0)^2}{4\sigma^{2}}
-\,\frac{\sigma^{2}(P-p_0)^2}{\hbar^{2}}
\right].\label{eq:init}
\end{equation}
In our numerical examples, we use the parameters $x_0=+1$, $p_0=-1.5$, $\sigma=0.5$ in units of $\hbar=1$.
In Fig.\ref{fig:init} we show the initial Husimi function calculated from the initial wave packet and a
cloud of initial classical trajectories distributed also according to the initial Husimi distribution.

\subsection{The harmonic dissipative Duffing Equation}

We begin our investigation with the harmonic case, where $\alpha>0$ and the nonlinearity is switched off, $\beta=0$.  This case can be solved analytically in both the classical and quantum cases. With $\beta=0$, the equation becomes the linear, damped, driven harmonic oscillator:
\begin{equation}
\ddot{x} + \delta \dot{x} + \alpha x = \gamma \cos(\omega t) \label{eq:main}
\end{equation}
The general solution $x(t)$ is the sum of the homogeneous solution $x_h(t)$ (the transient part) and a particular solution $x_p(t)$ (the steady-state part).
\[
x(t) = x_h(t) + x_p(t)
\]
First, we solve the homogeneous equation, which describes the natural, un-driven motion of the system:
\[
\ddot{x}_h + \delta \dot{x}_h + \alpha x_h = 0
\]
We assume a solution of the form $x_h(t) = e^{rt}$, which leads to the characteristic equation:
\[
r^2 + \delta r + \alpha = 0
\]
The roots of this quadratic equation are:
\[
r_{1,2} = -\frac{\delta}{2} \pm \sqrt{\left(\frac{\delta}{2}\right)^2 - \alpha}
\]
Let's define the natural undamped frequency as $\omega_0 = \sqrt{\alpha}$. 
Because $\delta > 0$, the term $e^{-(\delta/2)t}$ ensures that $\lim_{t \to \infty} x_h(t) = 0$. For this reason, $x_h(t)$ is called the {\em transient solution}. 
We don't develop this further; instead, we concentrate on the particular solution. It can be written in a compact form as a single cosine function with a phase shift $\phi$:
\[
x_p(t) = A \cos(\omega t - \phi)
\]
where the amplitude $A$ and phase angle $\phi$ are:
\begin{equation}
A = \frac{\gamma}{\sqrt{(\omega_0^2 - \omega^2)^2 + (\delta \omega)^2}},
\end{equation}
\begin{equation}
\tan(\phi) = \frac{\delta \omega}{\omega_0^2 - \omega^2}.
\end{equation}
This solution persists indefinitely as long as the driving force is applied and is known as the {\em steady-state solution}.

Starting an initial Gaussian cloud of classical trajectories in phase space, the transient part of the trajectories dies out over the timescale $\sim 1/\delta$ and the Gaussian concentrates around the steady state solution. In Fig.\ref{fig4:sub1} we show a cloud of classical trajectories that started from an initial Gaussian distribution (\ref{eq:init}) shown in Fig.\ref{fig:init} and converged to an attractor after many cycle periods $T_{cy}=2\pi/\omega$ of the external forcing.  The quantum mechanical counterpart of this classical system started from a Gaussian wave packet, whose Husimi distribution (also Gaussian) coincided with the classical density of trajectories.
In Fig.\ref {fig4:sub2}, we show the time-evolved Husimi distribution on a logarithmic scale after the same number of cycles. Finally, in Fig.\ref{fig4:sub3}, we superimpose the two figures to show that the Husimi distribution follows closely the classical distribution.

\begin{figure*}[t] 
    \centering
    \begin{subfigure}{0.32\textwidth}
        \includegraphics[width=\linewidth]{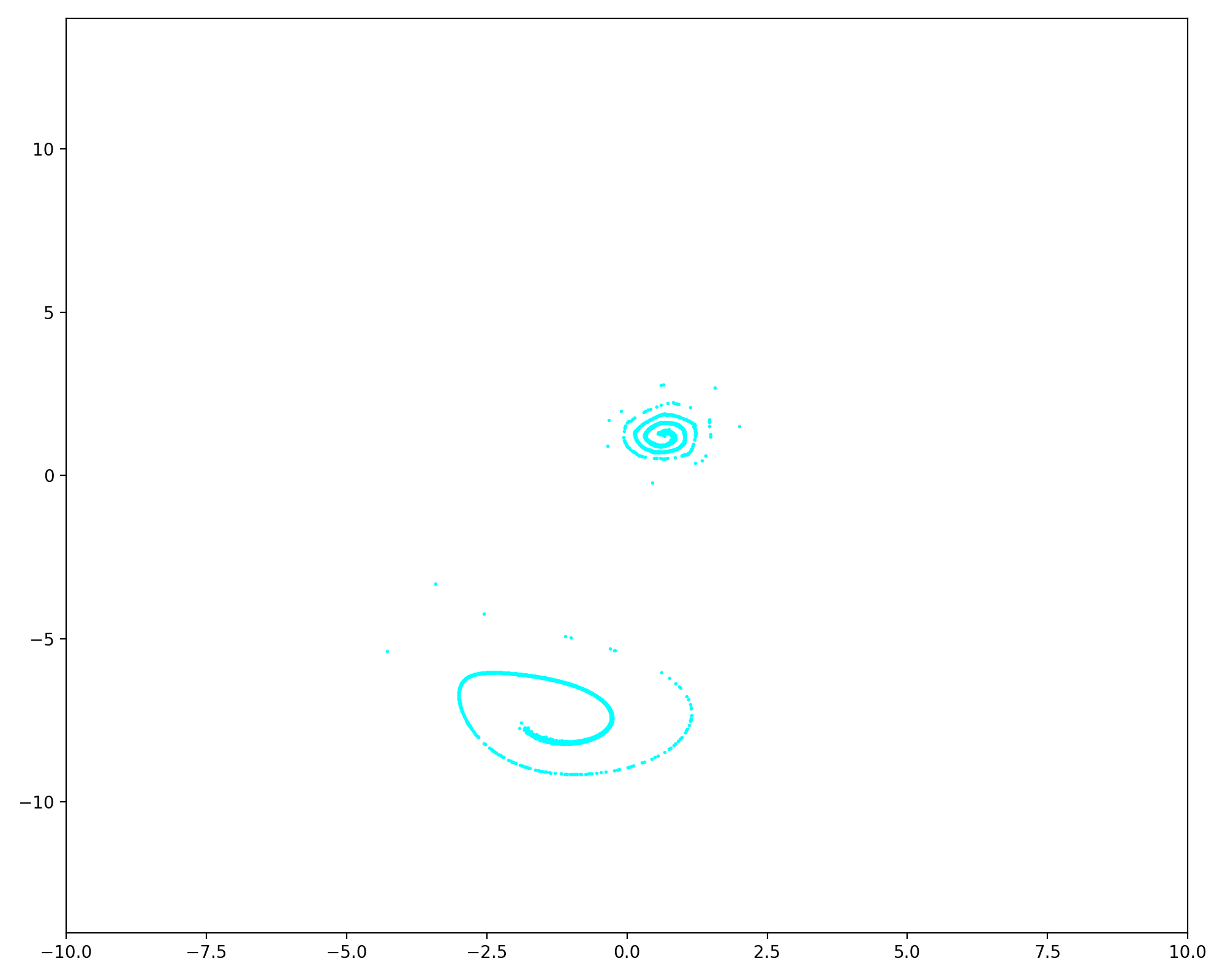}
        \caption{Poincaré return map $(x,P)$}
        \label{fig3:sub1}
    \end{subfigure}%
    \hfill 
    \begin{subfigure}{0.32\textwidth}
        \includegraphics[width=\linewidth]{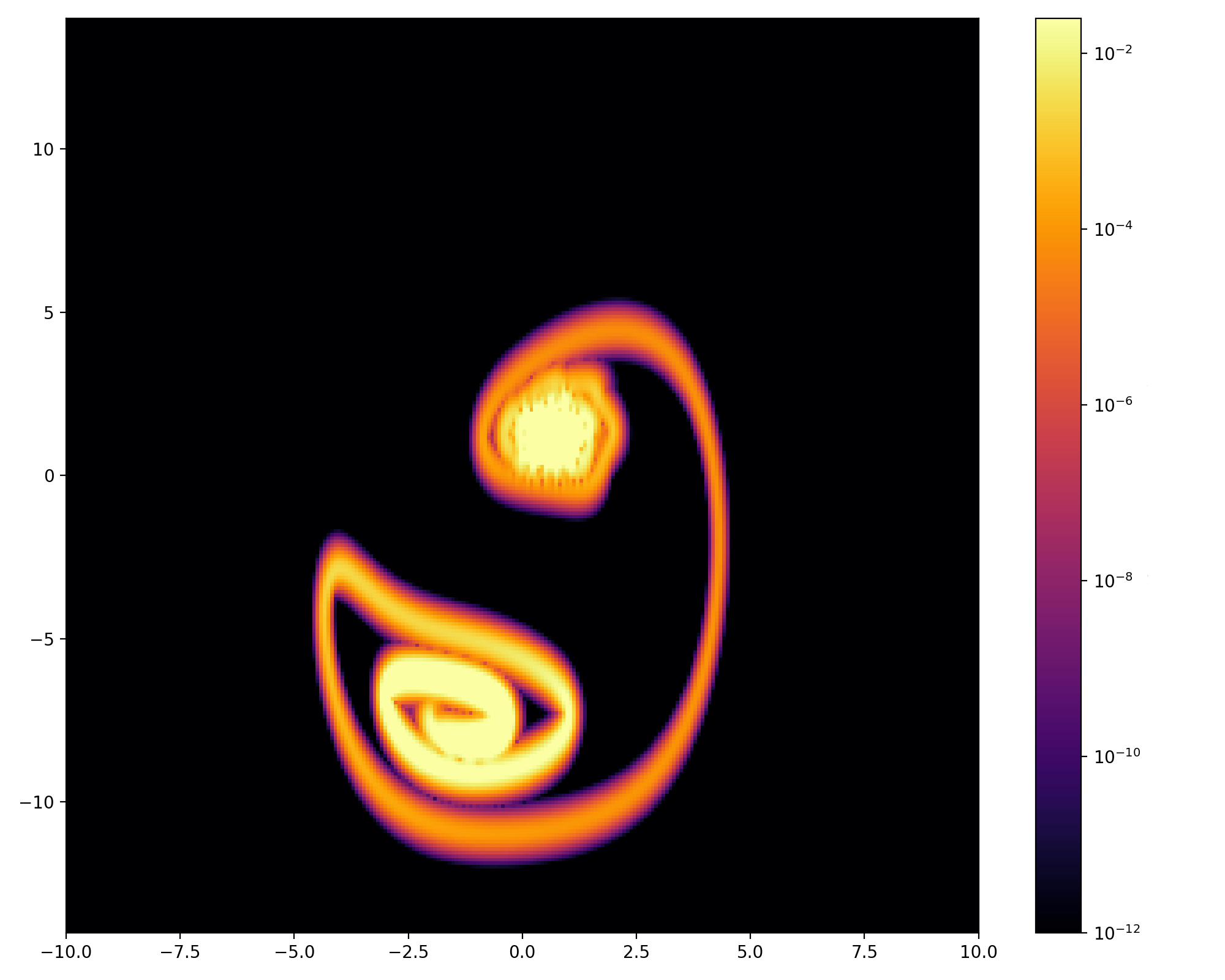}
        \caption{Husimi distribution $\log Q(x,P)$}
        \label{fig3:sub2}
    \end{subfigure}%
    \hfill 
    \begin{subfigure}{0.32\textwidth}
        \includegraphics[width=\linewidth]{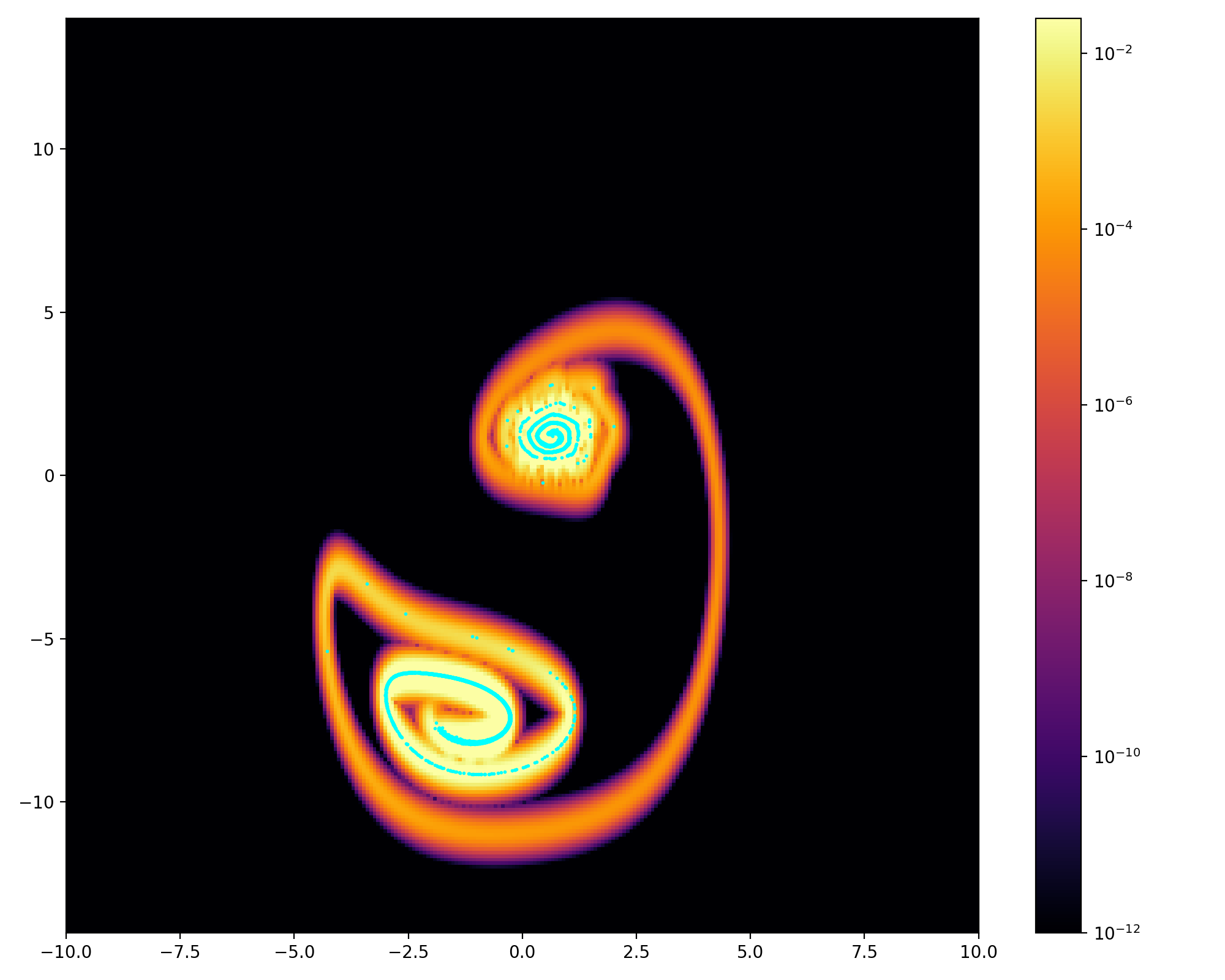}
        \caption{Poincaré + Husimi superimposed }
        \label{fig3:sub3}
    \end{subfigure}
    
    \caption{The non-chaotic dissipative Duffing equation $\alpha=+1$, $\beta=0.25$, $\delta=0.1$, $\gamma=2.5$, $\omega=2$. Snapshot of the classical and quantum evolution after $13.37$ cycles $T_{cy}$ of the external forcing.}
    \label{fig3:main}
\end{figure*}

\subsection{Transient dissipative Duffing Equation}

Next, we consider a somewhat more involved case, where we add nonlinearity to the harmonic case such that $\alpha>0$ and $\beta>0$. This situation is called the hardening spring oscillator. In this case, depending on the value of $\beta$, we can
have one stable steady state if $\beta<\beta_c$ or we can have two stable and one unstable solutions if $\beta>\beta_c$, depending on the frequency of driving $\omega$. This can lead to hysteresis in the frequency response of the amplitude, a phenomenon that has been extensively studied in the literature. The frequency response curve can be calculated in the following steps:
Assume a steady-state solution of the form
\begin{equation}
    x(t) \approx A \cos(\omega t - \phi),
\end{equation}
where $A$ is the amplitude and $\phi$ the phase lag. Using the identity
\[
    \cos^{3}(\omega t - \phi) = \tfrac{1}{4}\left(3\cos(\omega t - \phi) + \cos(3\omega t - 3\phi)\right),
\]
we approximate the cubic term by its first harmonic contribution:
\[
    x^{3}(t) \approx \tfrac{3}{4}A^{3}\cos(\omega t - \phi).
\]
Substituting into the governing equation and balancing first-harmonic terms yields
\begin{equation}
    \bigl(\alpha + \tfrac{3}{4}\beta A^{2} - \omega^{2}\bigr)A\cos\phi
    + (2\delta\omega)A\sin\phi = \gamma,
\end{equation}
\begin{equation}
    \bigl(\alpha + \tfrac{3}{4}\beta A^{2} - \omega^{2}\bigr)A\sin\phi
    - (2\delta\omega)A\cos\phi = 0.
\end{equation}
Eliminating $\phi$ gives the nonlinear algebraic equation for the amplitude:
\begin{equation}
    \Biggl[\Bigl(\alpha + \tfrac{3}{4}\beta A^{2} - \omega^{2}\Bigr)^{2}
           + (2\delta \omega)^{2}\Biggr] A^{2} = \gamma^{2}.
\end{equation}
This implicit relation between $A$ and $\omega$ defines the
{\em frequency–response curve} of the Duffing equation.

\begin{figure}[h]
    \centering
    \includegraphics[width=0.9\linewidth]{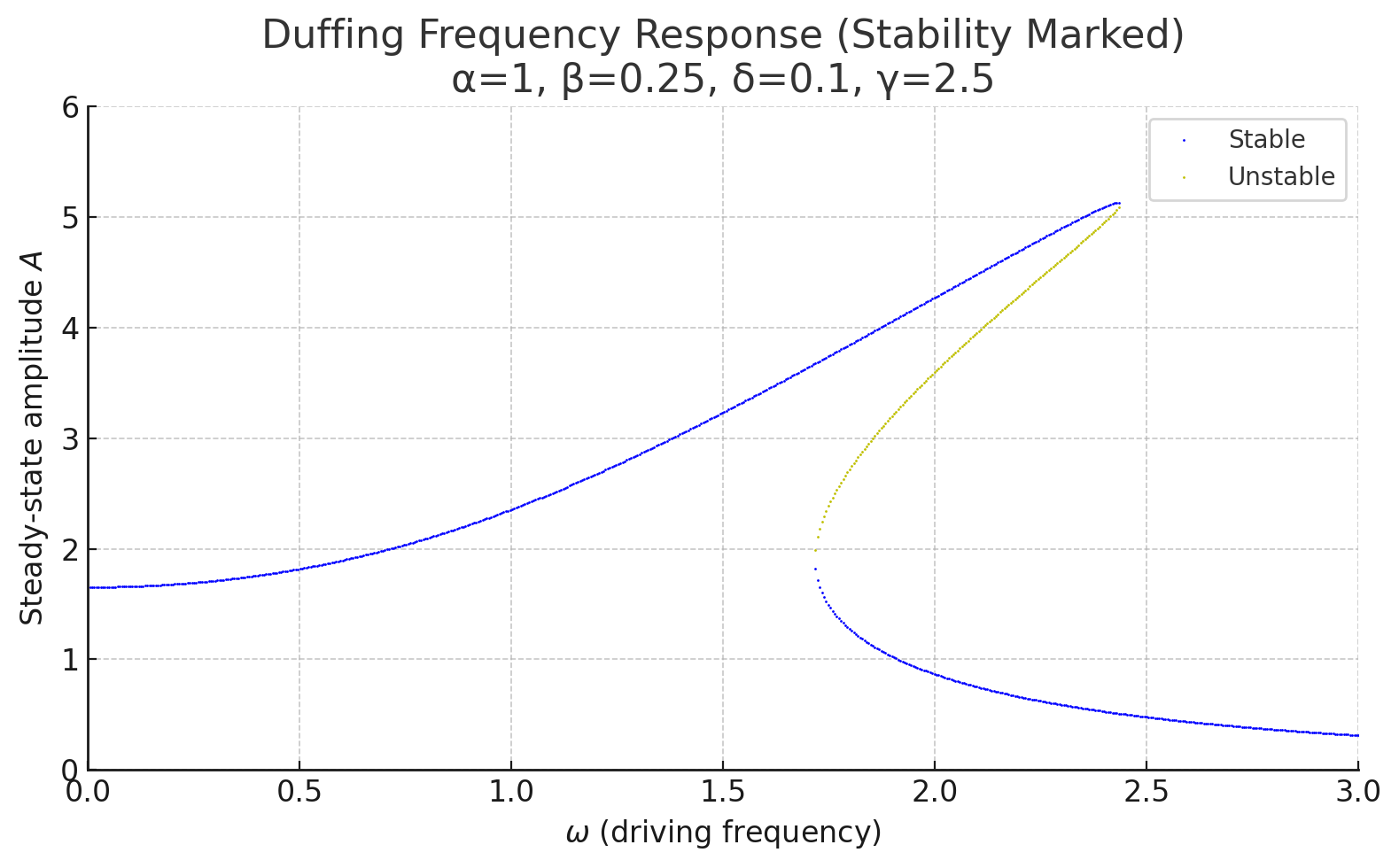}
    \caption{Frequency response curve of the Duffing Equation at parameters of Fig. \ref{fig3:main}. At frequency $\omega=2.0$, we have two stable and one unstable solution corresponding to two stable and one unstable periodic orbits in the phase space.}
    \label{fig:frequencyresponse}
\end{figure}

The frequency response curve of the system investigated in Fig. \ref{fig3:main} is shown in Fig. \ref{fig:frequencyresponse}. At the driving frequency $\omega=2.0$ two stable and one unstable solutions exist.  In Fig.\ref{fig3:main}, we present a Starting from our initial Gaussian cloud of classical initial conditions, some of the
trajectories quickly spiral to one of the stable solutions, which is an attractive focus. The rest of the trajectories spiral out of the unstable repulsive focus solution. The two spirals are clearly visible in Fig.\ref{fig3:sub1}. The classical dynamics is reflected in the Husimi distribution in Fig.\ref{fig3:sub2} and in the superimposed classical and quantum distributions in Fig.\ref{fig3:sub3}. 

\subsection{The conservative Duffing equation}

\begin{figure*}[t] 
    \centering
    \begin{subfigure}{0.32\textwidth}
        \includegraphics[width=\linewidth]{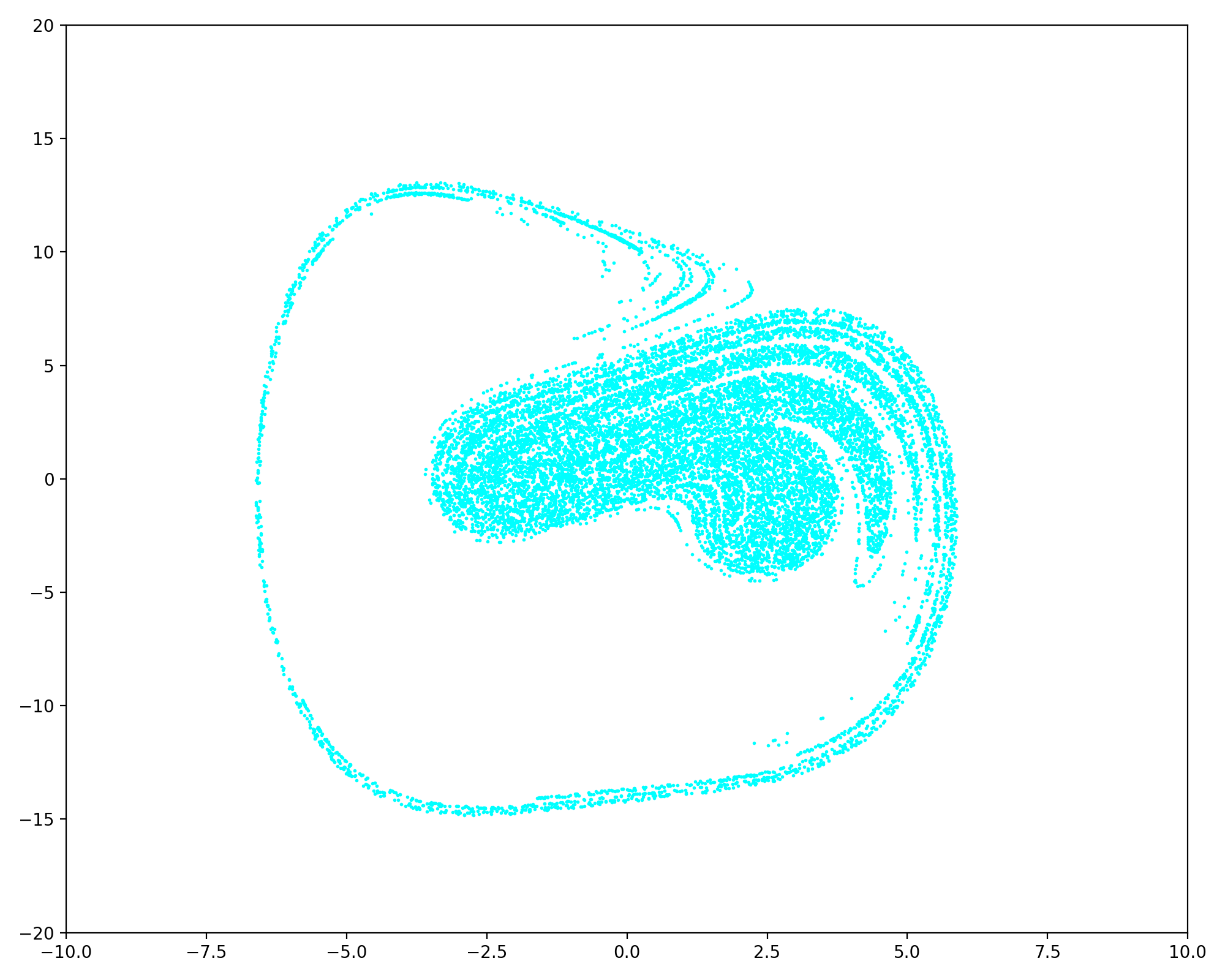}
        \caption{Poincaré return map $(x,P)$}
        \label{fig1:sub1}
    \end{subfigure}%
    \hfill 
    \begin{subfigure}{0.32\textwidth}
        \includegraphics[width=\linewidth]{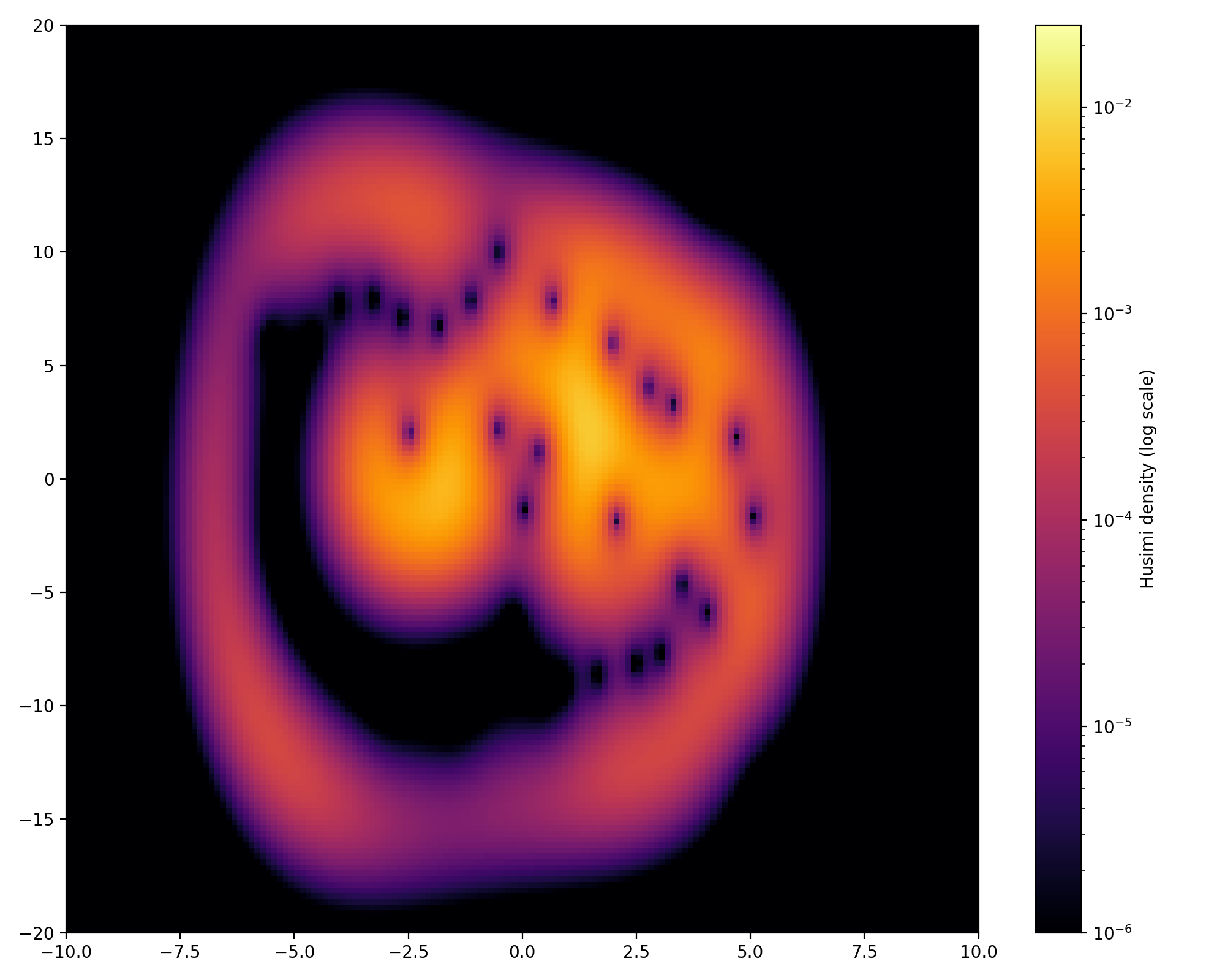}
        \caption{Husimi distribution $\log Q(x,P)$}
        \label{fig1:sub2}
    \end{subfigure}%
    \hfill 
    \begin{subfigure}{0.32\textwidth}
        \includegraphics[width=\linewidth]{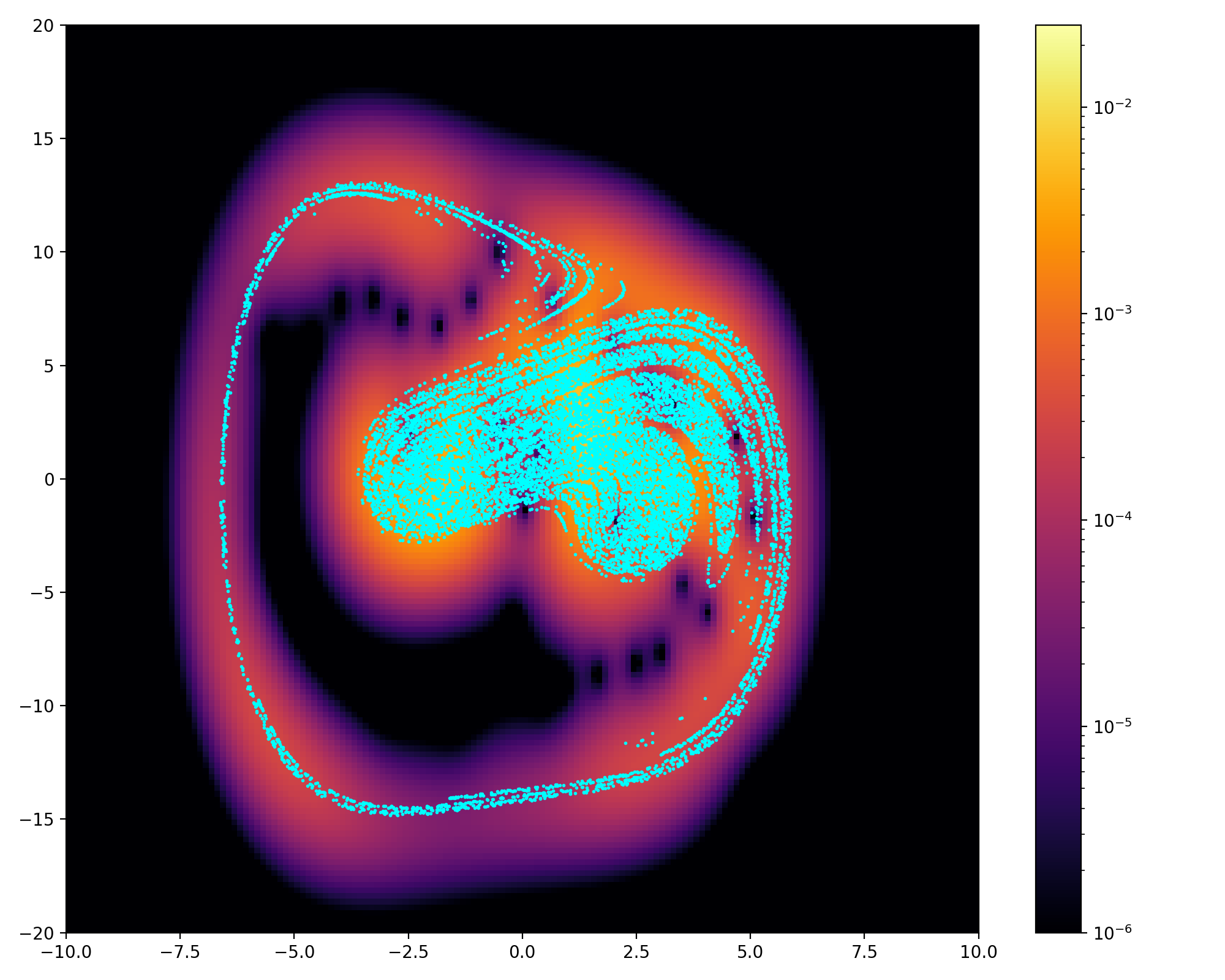}
        \caption{Poincaré + Husimi superimposed }
        \label{fig1:sub3}
    \end{subfigure}
    
    \caption{The conservative Duffing Equation $\alpha=-1.0$, $\beta=0.25$, $\delta=0.0$, $\gamma=2.5$, $\omega=2.0$. Snapshot of the classical and quantum evolution after $13.37$ cycles $T_{cy}$ of the external forcing.}
    \label{fig1:main}
\end{figure*}

The Poincaré section is a powerful tool for analyzing the dynamics of the conservative Duffing equation, $\ddot{x} + \alpha x + \beta x^3 = \gamma \cos(\omega t)$.  It reveals the underlying structure of the system's dynamics. For regular, quasi-periodic motion, the points on the section form well-defined, closed curves known as Kolmogorov-Arnold-Moser (KAM) tori. In contrast, chaotic trajectories generate a scatter of points that fill a region of the phase plane in a fractal-like pattern, often referred to as a "chaotic sea". A characteristic feature of the non-dissipative Duffing oscillator's Poincaré section is the coexistence of distinct regions: islands of stability, containing regular orbits (KAM tori), are surrounded by a larger chaotic sea, illustrating the complex interplay between order and chaos in this classic Hamiltonian system.
In the intricate dance between the quantum world and classical chaos, the enigmatic "Husimi zeros" have emerged as a powerful tool for understanding the quantum signatures of chaotic systems. These mathematical entities, which are the points where the Husimi function of a quantum state vanishes, provide a unique window into the underlying dynamics of a system, revealing a stark distinction between regular and chaotic behavior.

At the heart of what we know about Husimi zeros lies their remarkable ability to visually represent the nature of a quantum state. For systems that exhibit regular, predictable motion, the Husimi zeros are not randomly scattered; instead, they arrange themselves in highly structured, often one-dimensional patterns, akin to beads on a string, tracing out the regular tori in the classical phase space\cite{korsch1997zeros}.
In stark contrast, for a chaotic system, the Husimi zeros undergo a dramatic transformation. They no longer confine themselves to simple curves but instead spread out and appear to pepper the entire accessible two-dimensional phase space randomly. This "gas" of zeros is a direct visual hallmark of quantum chaos, indicating that the quantum state is exploring the full extent of its chaotic environment. The density of these zeros is fundamentally linked to Planck's constant, with approximately one zero occupying a phase-space area\cite{arranz2013onset} of $2\pi\hbar$. While a direct and simple mathematical formula linking the distribution of Husimi zeros to classical measures of chaos, such as Lyapunov exponents, remains an area of active research, the qualitative connection is clear: the more chaotic the system (as quantified by positive Lyapunov exponents), the more uniformly the Husimi zeros will be distributed throughout the phase space.

In Fig.\ref{fig1:main} we show a conservative Duffing Equation at our standard parameters. In Fig.\ref{fig1:sub1} we show the cloud of classical trajectories after $13.37$ cycles. 
The KAM tori structure and the diffusion in the chaotic sea are visible. The corresponding Husimi distribution in
Fig.\ref{fig1:sub2} possesses zeros (black dots) indicating the existence of quantum chaos in those areas. In the superimposed Fig. \ref{fig1:sub3}, we can see that the chaotic sea in the Poincaré section and the area with irregular patterns of Husimi zeros strongly overlap.
The regular lines of Husimi zeros separate the chaotic sea from areas that are impermeable to chaotic motion due to unbroken KAM tori bordering them.

\subsection{The chaotic dissipative Duffing equation}
\begin{figure*}[t] 
    \centering
    \begin{subfigure}{0.32\textwidth}
        \includegraphics[width=\linewidth]{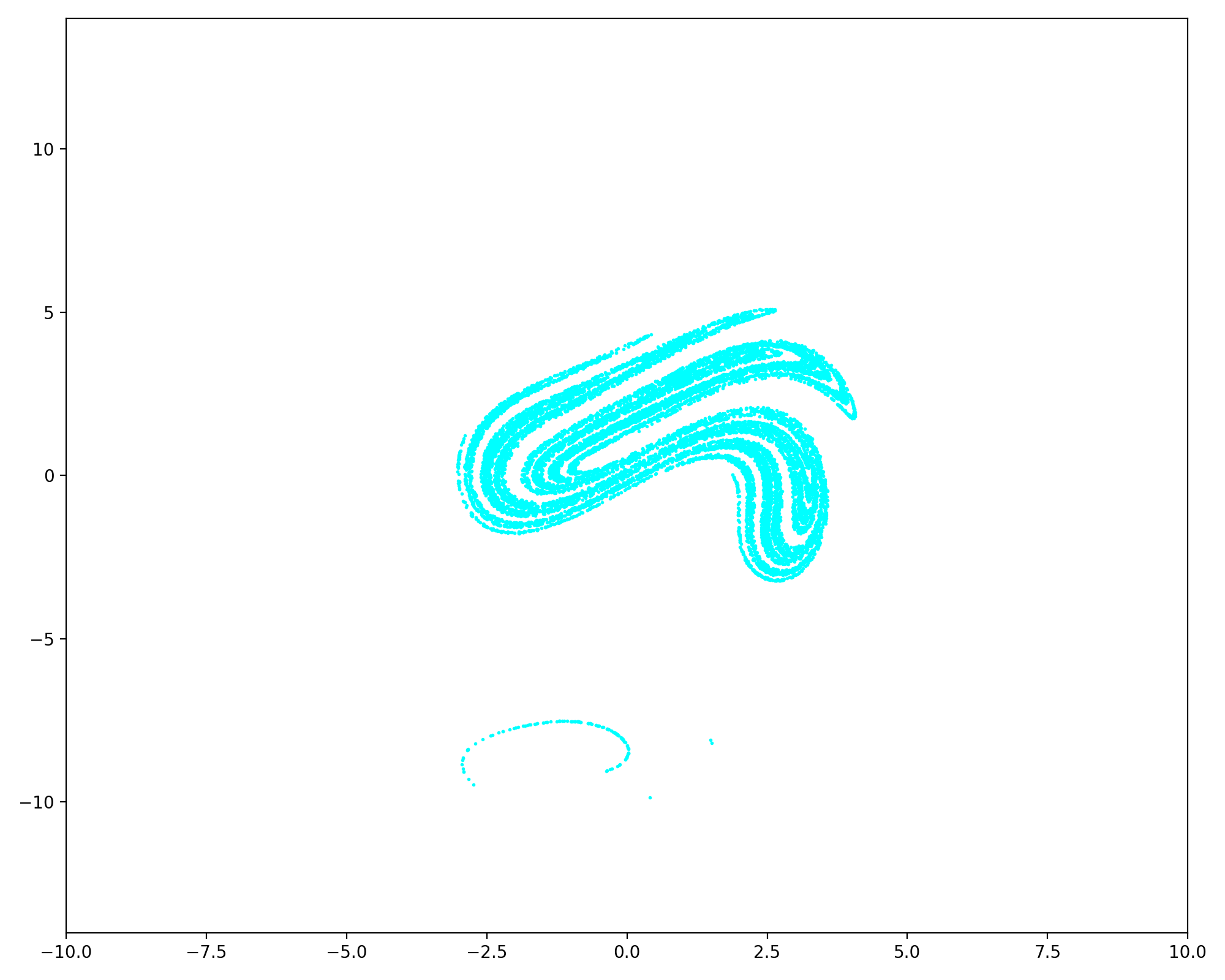}
        \caption{Poincaré return map $(x,P)$}
        \label{fig2:sub1}
    \end{subfigure}%
    \hfill 
    \begin{subfigure}{0.32\textwidth}
        \includegraphics[width=\linewidth]{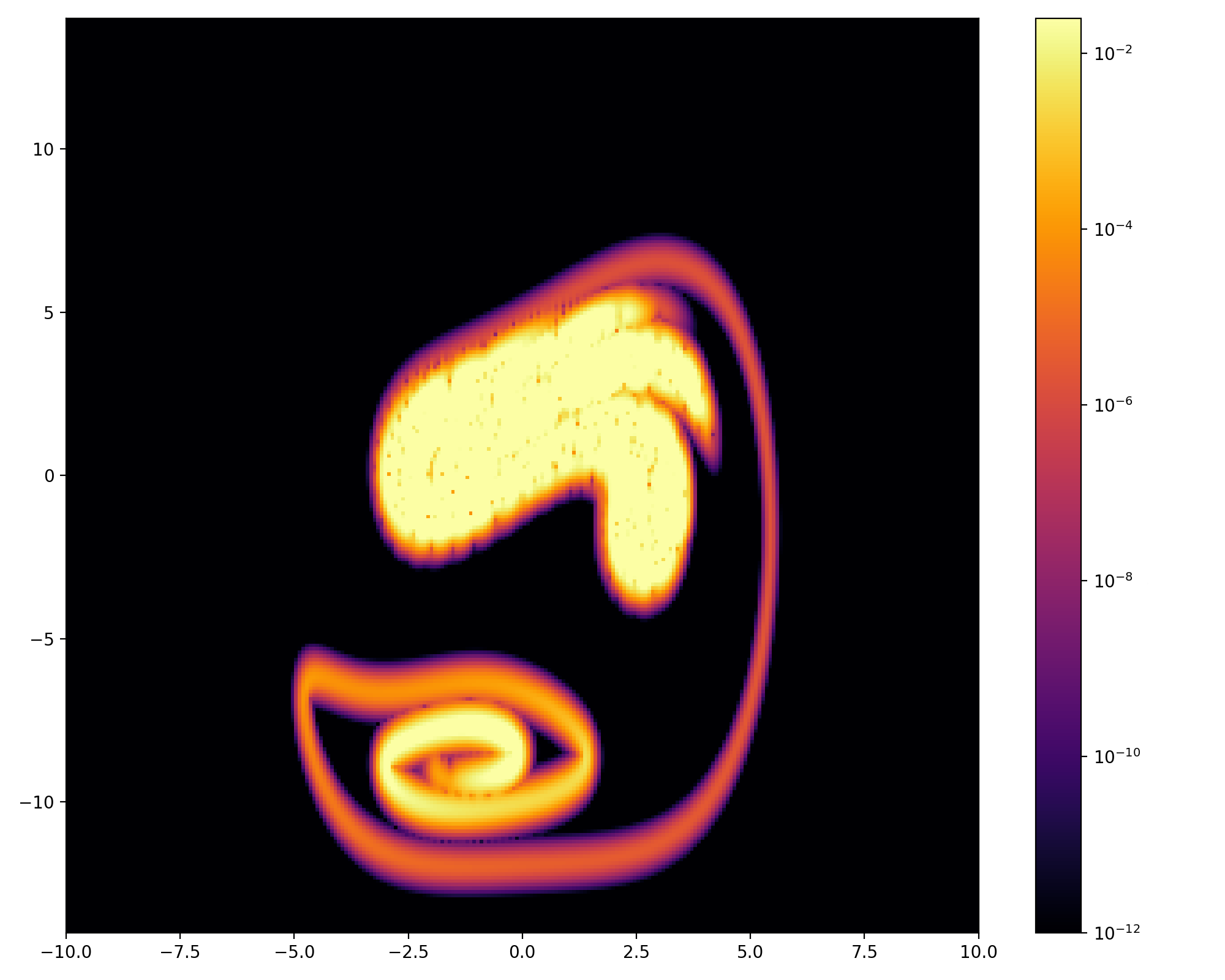}
        \caption{Husimi distribution $\log Q(x,P)$ }
        \label{fig2:sub2}
    \end{subfigure}%
    \hfill 
    \begin{subfigure}{0.32\textwidth}
        \includegraphics[width=\linewidth]{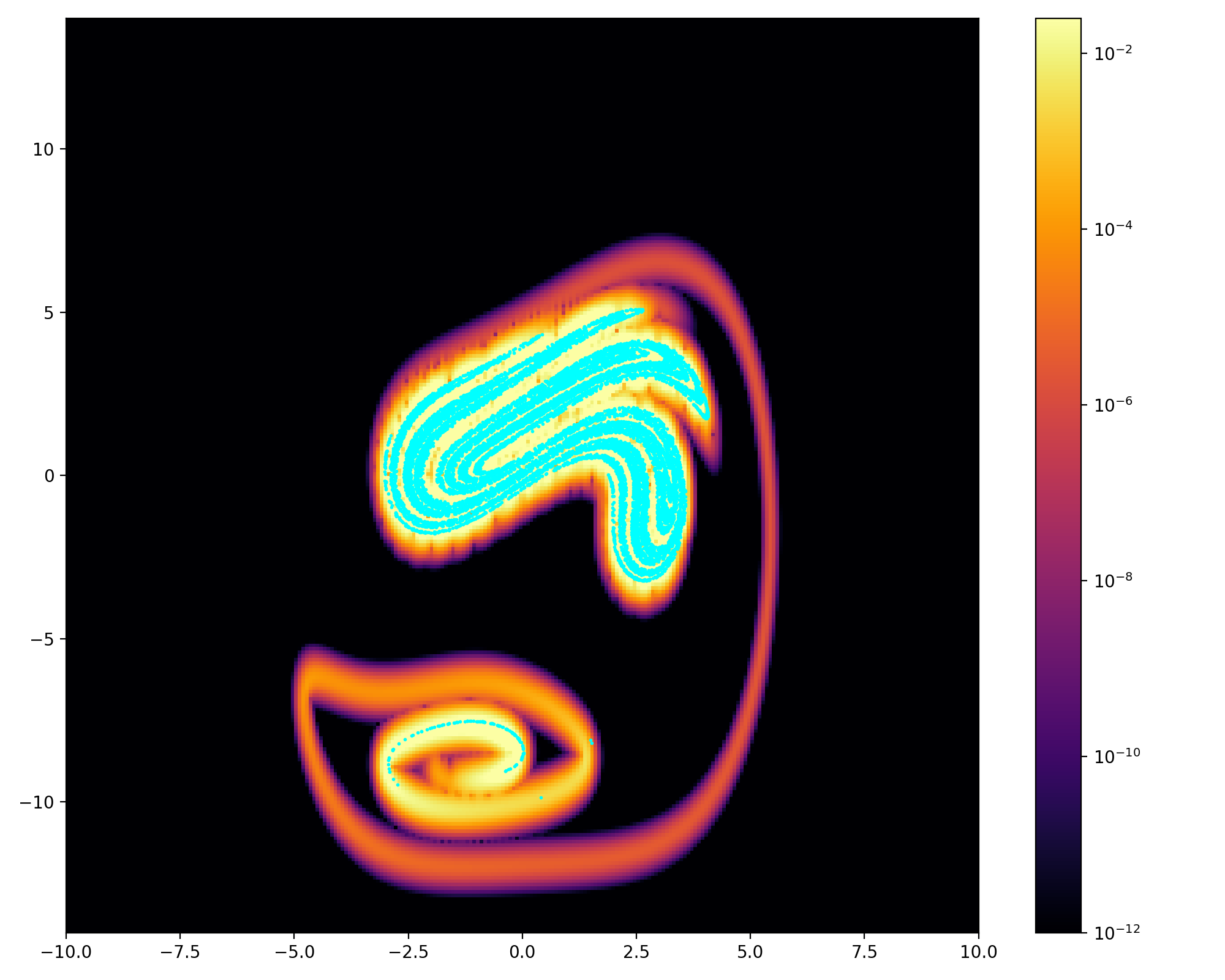}
        \caption{Poincaré + Husimi superimposed }
        \label{fig2:sub3}
    \end{subfigure}
    
    \caption{The chaotic dissipative Duffing equation $\alpha=-1.0$, $\beta=0.25$, $\delta=0.1$, $\gamma=2.5$, $\omega=2.0$. Snapshot of the classical and quantum evolution after $13.37$ cycles $T_{cy}$ of the external forcing.}
    \label{fig2:main}
\end{figure*}

Finally, we arrive at the central result of our investigation: an example of the dissipative Duffing equation exhibiting a strange attractor, with parameters set to $\alpha=-1.0$, $\beta=0.25$, $\delta=0.1$, $\gamma=2.5$, and $\omega=2.0$. This section presents, for the first time, a direct visualization of the quantum mechanical counterpart to a classical strange attractor within the Caldirola-Kanai framework, representing the culmination of our analysis.

The classical dynamics, shown in the Poincaré map in Fig.\ref{fig2:sub1}, reveal the beautiful and complex structure of a strange attractor. This object is the result of two competing mechanisms: the repeated stretching and folding of trajectories due to the nonlinear dynamics, which generates sensitive dependence on initial conditions (chaos), and the uniform contraction of phase-space volume due to dissipation ($\delta > 0$). This contraction ensures that trajectories are confined to a bounded region, and the combination of these effects leads to an attractor with an intricate, self-similar fractal structure.

The corresponding quantum state is revealed through the Husimi distribution in Fig.\ref{fig2:sub2}. This distribution is the long-time limit of an initially simple Gaussian wave packet that has been sculpted by the competing forces of nonlinear driving and damping. The resulting quantum probability density is strikingly localized in the phase space.

The profound connection between the classical and quantum realms is made explicit in Fig.\ref{fig2:sub3}, which superimposes the classical Poincaré points onto the quantum Husimi distribution. Several key observations can be made:

{\em Quantum Localization on the Attractor:} The Husimi distribution's high-probability regions meticulously trace the backbone of the classical strange attractor. This stands in stark contrast to the conservative chaotic case (Fig.\ref{fig1:main}), where the quantum state delocalizes into a "chaotic sea" populated by a gas of Husimi zeros. Here, dissipation acts as a powerful localizing force, preventing the wave function from exploring the entire phase space and instead compelling it to collapse onto the lower-dimensional manifold of the attractor.

{\em Smoothing of Fractal Structure:} While the classical attractor possesses structure on an arbitrarily fine scale, the Husimi distribution is inherently smooth. This ``quantum smoothing'' is a direct manifestation of the uncertainty principle, encapsulated in our formalism by the time-dependent effective Planck's constant, $\hbar(t)$. The finite width of the coherent states used to construct the Husimi function sets a fundamental limit on phase-space resolution, blurring out the classical fractal's finest details.

{\em A New Dynamical State:} This localized, complex distribution is fundamentally different from the other cases we examined. It is not the simple, stable fixed point of the harmonic system (Fig.\ref{fig4:main}), nor is it the transient state spiraling toward two distinct attractors seen in the non-chaotic hardening spring (Fig.\ref{fig3:main}). It is a uniquely quantum signature of a system that is simultaneously chaotic, dissipative, and in a steady state.

In summary, Fig.\ref{fig2:main} provides a compelling visual narrative of how the defining features of classical dissipative chaos manifest in a quantum system. The Husimi distribution serves as a quantum photograph of the strange attractor, faithfully capturing its global geometry while smoothing over its classical fractal intricacies. This visualization validates the Caldirola-Kanai approach as a powerful tool for exploring the quantum-classical transition in open, chaotic systems, providing a complementary geometric viewpoint to more familiar Lindblad- and Langevin-based analyses.

\section{Semiclassification ofout-of-time-ordered correlator}

The  OTOC (\ref{eq:otoc}) serves as a key diagnostic tool for identifying quantum chaos, drawing a parallel to the classical notion of extreme sensitivity to initial conditions. In a quantum system, the OTOC measures the extent to which two initially commuting operators fail to commute after one is evolved in time. For a chaotic quantum system, this non-commutativity grows exponentially at early times, a behavior quantified by a quantum Lyapunov exponent. This exponential growth signifies the rapid scrambling of quantum information throughout the system's degrees of freedom, a hallmark of quantum chaos. In contrast, for non-chaotic (integrable) systems, the OTOC typically exhibits a much slower, often polynomial, growth or oscillatory behavior.

Extracting the classical Lyapunov exponent, a key signature of chaos, from the short-time behavior of OTOC in quantum systems is a subtle and often difficult task. While in theory, the OTOC for a chaotic system is expected to exhibit exponential growth at early times (\ref{eq:lambdaq}).
This behavior is often preceded and obscured by other effects, making a clean extraction challenging.

One of the primary difficulties is that the exponential growth phase does not start immediately at $t=0$. There is an initial period during which the OTOC's behavior is non-universal and system-dependent\cite{fortes2019gauging}. During this initial phase, the correlator's growth is typically polynomial (often quadratic) rather than exponential. This initial non-exponential growth is a quantum effect that has no classical counterpart and can be mistaken for the onset of chaos if not carefully distinguished from it. The true exponential growth, characteristic of chaos, only emerges after this initial transient period. 

\begin{figure}[h!]
    \centering
    \includegraphics[width=0.9\linewidth]{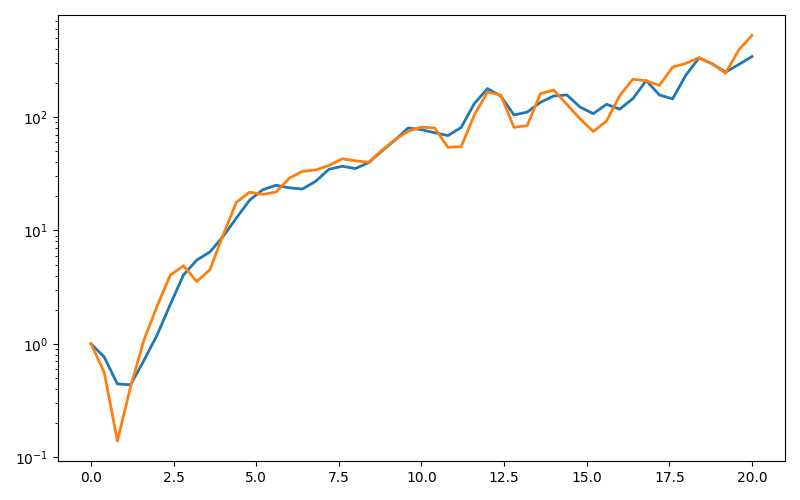}
    \caption{Time evolution of OTOC (\ref{eq:otoc}) for $\alpha=-1.0$, $\beta=0.25$, $\gamma=2.5$, $\omega=2.0$ for the conservative chaos $\delta=0.0$ (orange) and for the dissipative chaos $\delta=0.1$ (blue) on a semi-logarithmic plot.}
    \label{fig:otoc}
\end{figure}

The window for observing this exponential growth is further constrained by the Ehrenfest time, $\tau_E$. This is the timescale at which quantum effects become significant and the correspondence between classical and quantum dynamics breaks down. The exponential growth of the OTOC, which signals classical-like chaotic behavior, is only expected to hold for times shorter than the Ehrenfest time. For times $t>\tau_E$, quantum interference effects take over, and the OTOC's growth slows down and eventually saturates\cite{tsuji2018bound}. In systems with a small separation of timescales between the initial non-universal behavior and the Ehrenfest time, the window for observing a clear exponential growth can be very narrow, or even non-existent.

\begin{figure*}[t] 
    \centering
    \begin{subfigure}{0.32\textwidth}
        \includegraphics[width=\linewidth]{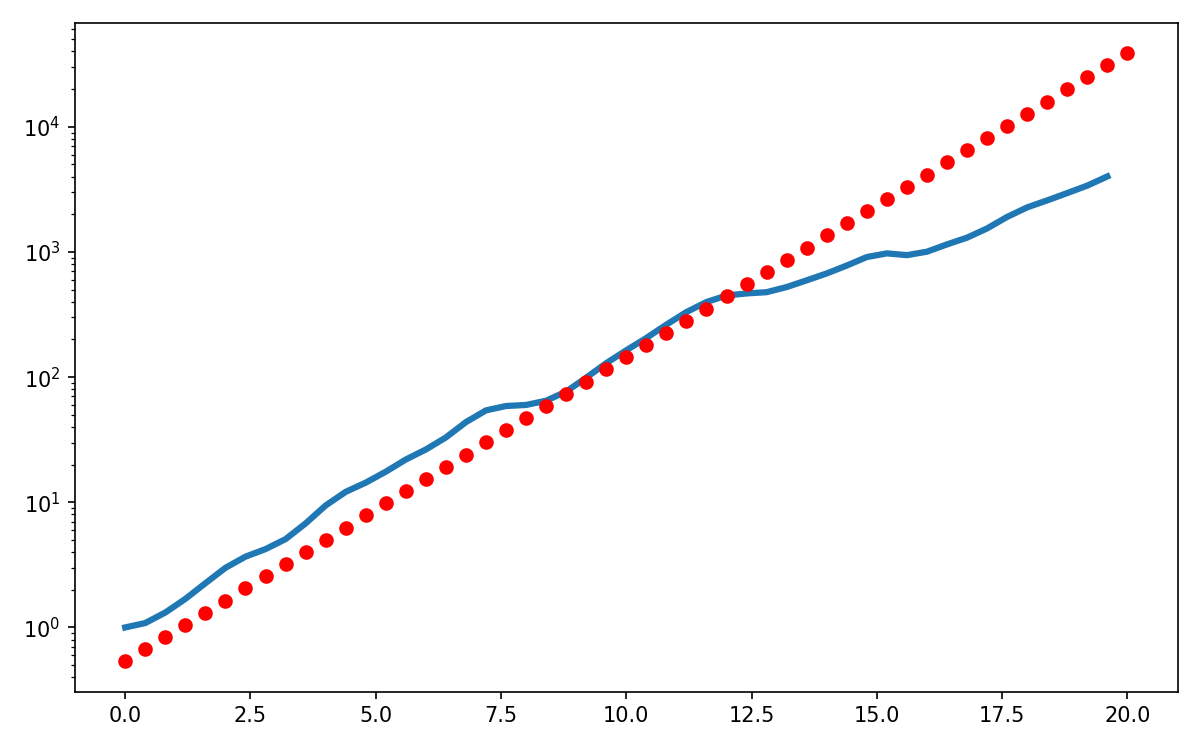}
        \caption{$\delta=0.2$ and $\lambda=0.2799$}
        \label{fig5:sub1}
    \end{subfigure}%
    \hfill 
    \begin{subfigure}{0.32\textwidth}
        \includegraphics[width=\linewidth]{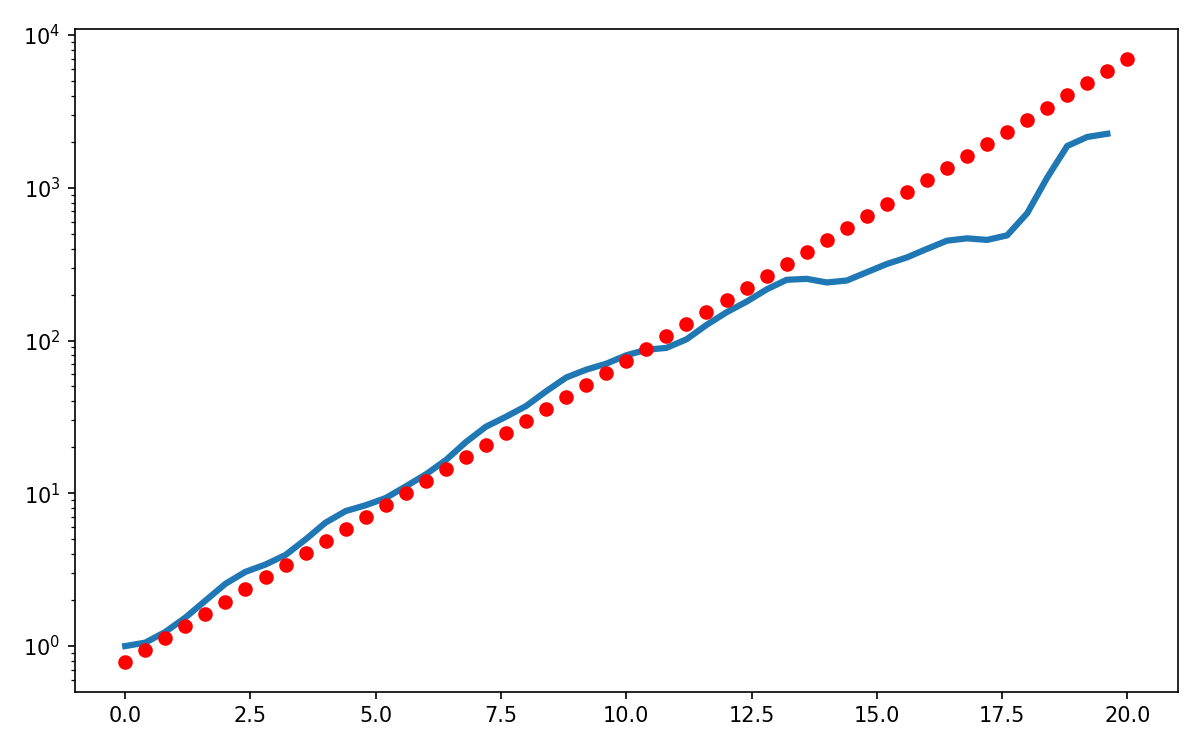}
        \caption{$\delta=0.3$ and $\lambda=0.2269$}
        \label{fig5:sub2}
    \end{subfigure}%
    \hfill 
    \begin{subfigure}{0.32\textwidth}
        \includegraphics[width=\linewidth]{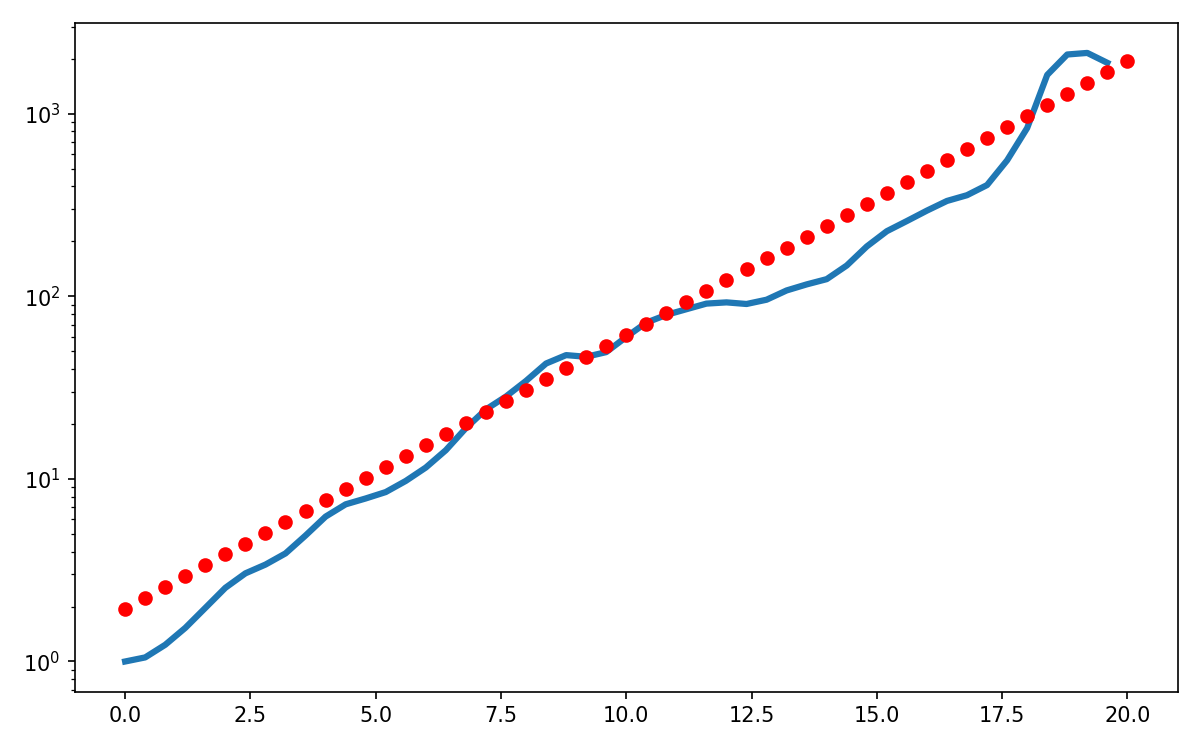}
        \caption{$\delta=0.35$ and $\lambda=0.1728$}
        \label{fig5:sub3}
    \end{subfigure}
    
    \caption{OTOC for the chaotic dissipative Duffing equation $\alpha=-1.0$, $\beta=0.25$, $\gamma=2.5$, $\omega=2.0$ on a semi-logarithmic plot. Red dots represent $const \times e^{2\lambda t}$, where $\lambda$ is the Lyapunov exponent of the classical system. The deviation from classical exponential behavior occurs later with an increase in the dissipation rate. }
    \label{fig5:main}
\end{figure*}

We calculated the OTOC numerically according to the definition (\ref{eq:otoc}) for the Duffing equation, using the parameters from the previous sections ($\alpha=-1.0$, $\beta=0.25$, $\gamma=2.5$, $\omega=2.0$). We initialized the system according to equation (\ref{eq:init}). First, we show the conservative ($\delta=0.0$) quantum chaotic and the moderately dissipative chaotic cases ($\delta=0.1$) in Fig.\ref{fig:otoc}. We can observe an initial transient behavior, followed by a very short, barely identifiable exponential period, and then saturation. The moderately dissipative case does not differ significantly from the conservative case in this respect. 

As we discussed in Section\ref{section:semi}, the rate of dissipation determines how quickly the system transitions from the quantum to the semiclassical behaviour. If we increase the rate of dissipation, the OTOC changes significantly. In Fig.\ref{fig5:main} we show the OTOC for $\delta=0.2, 0.3$ and $0.35$. The initial part becomes exponential, and we found that the exponent is in good agreement with the classical Lyapunov exponent $\lambda_Q\approx\lambda$. The Lyapunov exponent has been calculated classically by solving 
equation (\ref{eq:stability}) for a long classical trajectory and then using equation (\ref{eq:lyapunov}). 

This result provides a numerical validation of the semiclassical interpretation of the Caldirola-Kanai framework. The enhanced dissipation rate accelerates the decay of the effective Planck's constant $\hbar(t)=\hbar e^{-\delta t}$, driving the system more rapidly toward the semiclassical limit. This rapid classicalization effectively suppresses the initial, purely quantum, non-universal growth phase of the OTOC and extends the time window before quantum saturation effects dominate. Consequently, a clearer exponential growth regime emerges, allowing for a direct and accurate extraction of the classical Lyapunov exponent. The OTOC, within this framework, thus serves not only as an indicator of quantum chaos but also as a dynamic probe of the quantum-to-classical transition, demonstrating how dissipation can restore a key signature of classical chaos in a quantum system.

\section{Conclusions}

In this work, we have successfully provided the first direct visualization of a quantum strange attractor within the Caldirola-Kanai (CK) framework. By investigating the quantum dynamics of the driven, dissipative Duffing oscillator, we have demonstrated how the interplay of chaotic evolution and energy loss sculpts an initially simple quantum state into a complex, localized structure in phase space.

Our central tool for this analysis was the Husimi distribution, adapted to the CK formalism's time-dependent effective Planck's constant. We systematically explored four distinct dynamical regimes, culminating in our main result. In the harmonic and non-chaotic dissipative cases, the quantum wave packet's evolution faithfully tracked the convergence of classical trajectories to simple attractors. In the conservative chaotic regime, we observed the expected delocalization of the Husimi distribution into a "chaotic sea," with the irregular pattern of Husimi zeros clearly marking the regions of classical chaos.

The key finding of this paper is the behavior in the chaotic dissipative regime. Here, unlike the conservative case, dissipation acts as a powerful localizing mechanism, compelling the quantum state to collapse onto the intricate, filamentary structure of the classical strange attractor. This "quantum photograph" of the attractor, however, is inherently smoothed by the uncertainty principle, blurring the infinitely fine fractal details of its classical counterpart. This result offers a compelling geometric picture of how a quantum system can inherit the global structure of classical chaos while respecting fundamental quantum limitations.

Our analysis of the out-of-time-ordered correlator (OTOC) provided a dynamic validation of the semiclassical nature of the CK model. We showed that extracting a clear exponential growth, which is characteristic of the classical Lyapunov exponent, is difficult in conservative or weakly dissipative systems due to quantum effects. However, by increasing the dissipation rate, the system is driven more rapidly towards the classical limit. This increase in dissipation suppresses the initial, non-universal quantum growth and extends the Ehrenfest time. As a result, we observe a clear exponential regime in the OTOC, where the exponent aligns remarkably well with the classical Lyapunov exponent.

In summary, this study validates the Caldirola-Kanai approach as a powerful and intuitive framework for exploring the quantum signatures of dissipative chaos. By offering a direct phase-space visualization, our work provides a complementary perspective to established open-system formalisms and sheds new light on the fundamental mechanisms governing the quantum-classical transition in complex, open systems.

\begin{acknowledgments}
This research was supported by the Ministry of Culture and Innovation and the National Research, Development and Innovation Office within the Quantum Information National Laboratory of Hungary (Grant No. 2022-2.1.1-NL-2022-00004).
While preparing this work, the authors used Grammarly (including its AI services) to improve grammar and text composition. After using this tool, the authors reviewed and edited the content as needed and took full responsibility for the publication's content.

\end{acknowledgments}

\section*{AUTHOR DECLARATIONS}
\noindent {\bf Conflict of Interest}\\
The authors have no conflicts to disclose.\\

\noindent {\bf Author Contributions}\\
{\bf Bence Dárdai and Gábor Vattay}: Conceptualization (equal); Formal analysis
(equal); Investigation (equal); Methodology (equal); Visualization (equal); Writing – original draft (equal); Writing – review \& editing (equal).\\

\section*{Data Availability Statement}

The data that support the findings of this study are available from the corresponding author upon reasonable request.

\nocite{*}
\bibliography{aipsamp}

\end{document}